\documentclass[aps,prd,twocolumn,
preprintnumbers,
showpacs,floatfix]{revtex4}

\usepackage{bm}
\usepackage{epsfig}
\usepackage{graphics}
\usepackage{psfrag}
\usepackage{amsmath}
\usepackage{textcomp}

\newcommand{\eeq}{\end{equation}}
\newcommand{\beq}{\begin{equation}}
\newcommand{\ba}{\begin{array}}
\newcommand{\ea}{\end{array}}
\newcommand{\bea}{\begin{eqnarray}}
\newcommand{\eea}{\end{eqnarray}}
\newcommand{\vev}[1]{\langle #1\rangle}

\newcommand{\eps}{\epsilon}
\newcommand{\veps}{\varepsilon}

\newcommand{\si}{\sigma}

\newcommand{\lsim}{\mbox{\raisebox{-.9ex}
{~$\stackrel{\mbox{$<$}}{\sim}$~}}}

\newcommand{\pb}{\bar{\phi}}
\newcommand{\Hb}{\bar{H}}
\newcommand{\ka}{\kappa}
\newcommand{\la}{\lambda}

\newcommand{\ten}[1]{\,\cdot 10^{#1}}
\newcommand{\units}[1]{\,\text{#1}}

\newcommand{\1}{\mathit{1}}
\newcommand{\m}{\mathit{m}}
\newcommand{\ti}{\mathit{t_1}}
\newcommand{\tii}{\mathit{t_2}}

\DeclareMathOperator{\tr}{tr}

\begin{document}

\preprint{UT--STPD--2/07}

\title{New smooth hybrid inflation}

\author{George Lazarides}
\email{lazaride@eng.auth.gr}
\author{Achilleas Vamvasakis}
\email{avamvasa@gen.auth.gr}
\affiliation{Physics Division, School of
Technology, Aristotle University of
Thessaloniki, Thessaloniki 54124, Greece}

\date{\today}

\begin{abstract}
\par
We consider the extension of the supersymmetric
Pati-Salam model which solves the $b$-quark
mass
problem of supersymmetric grand unified models
with exact Yukawa unification and universal
boundary conditions and leads to
the so-called new shifted hybrid inflationary
scenario. We show that this model can also lead
to a new version of smooth hybrid inflation based
only on renormalizable interactions provided that
a particular parameter of its superpotential is
somewhat small. The potential possesses valleys
of minima with classical inclination, which can
be used as inflationary paths. The model is
consistent with the fitting of the three-year
Wilkinson microwave anisotropy probe data by
the standard power-law cosmological model with
cold dark matter and a cosmological constant. In
particular, the spectral index turns out to be
adequately small so that it is compatible with
the data. Moreover, the Pati-Salam gauge group is
broken to the standard model gauge group during
inflation and, thus, no monopoles are formed at
the end of inflation. Supergravity corrections
based on a non-minimal K\"{a}hler potential with
a convenient choice of a sign keep the spectral
index comfortably within the allowed range
without generating maxima and minima of the
potential on the inflationary path. So, unnatural
restrictions on the initial conditions for
inflation can be avoided.

\end{abstract}

\pacs{98.80.Cq}
\maketitle

\section{Introduction}
\label{sec:intro}

\par
In recent years, a plethora of precise
cosmological observations on the cosmic microwave
background radiation and the large-scale
structure in the universe has strongly favored
the idea of inflation \cite{inflation} (for a
review see e.g. Ref.~\cite{lectures}). Therefore,
the construction of realistic models of inflation
which are based on particle theory and are
consistent with all cosmological and
phenomenological requirements is an important
task. One of the most promising inflationary
models is, undoubtedly, the well-known hybrid
inflation \cite{linde}. This scenario is
\cite{cop,dss} naturally realized in the context
of supersymmetric (SUSY) grand unified theory
(GUT) models based on gauge groups with rank
greater than or equal to five.

\par
An attractive rank five gauge group is certainly
the Pati-Salam (PS) group $G_{\rm PS}=
{\rm SU}(4)_c\times{\rm SU}(2)_{\rm L}\times
{\rm SU}(2)_{\rm R}$ \cite{pati}. This is the
simplest GUT gauge group
which can lead \cite{hw} to ``asymptotic'' Yukawa
unification \cite{als}, i.e. the exact equality of
the third generation Yukawa coupling constants at
the GUT scale. Moreover, SUSY PS GUT models are
motivated \cite{tye} (see also Ref.~\cite{kane})
from the recent D-brane set-ups and can also arise
\cite{leontaris} from the standard weakly coupled
heterotic string.

\par
The standard realization of the SUSY hybrid
inflation scenario is based on a renormalizable
superpotential. In this model, the spontaneous
breaking of the GUT gauge symmetry takes place at
the end of inflation and, thus, topological
defects are copiously formed \cite{smooth} if
they are predicted by this symmetry breaking. The
spontaneous breaking of $G_{\rm PS}$ to the
standard model (SM) gauge group $G_{\rm SM}$ does
predict topologically stable magnetic monopoles,
which carry \cite{magg} two units of Dirac
magnetic charge. So, these monopoles are
overproduced \cite{shift} at the end of standard
SUSY hybrid inflation leading to a cosmological
disaster.

\par
A possible solution to this problem may be
obtained \cite{shift,smooth,talks} by including
into the standard superpotential for hybrid
inflation the leading non-renormalizable term,
which cannot be excluded by any symmetry and can
be comparable with the trilinear term of the
standard superpotential. Actually, we have two
options. We can either keep \cite{shift} both
these terms or remove \cite{smooth} the trilinear
term by imposing an appropriate discrete symmetry
and keep only the leading non-renormalizable term.
In the former case, there appears a new
``shifted'' classically flat valley of local
minima. This valley acquires a slope at the
one-loop level and can be used as an alternative
inflationary path.
The resulting scenario is known as shifted hybrid
inflation \cite{shift}. The latter option leads
to the existence of an inflationary path which
possesses an inclination already at the classical
level. In contrast to the standard and shifted
hybrid inflation scenarios where inflation
terminates abruptly and is followed by a
``waterfall'' regime, in this case, it ends
smoothly by saturating the slow-roll conditions.
So, the name smooth hybrid inflation was coined
\cite{smooth} for this scenario. In both shifted
and smooth hybrid inflation, the GUT gauge group
$G_{\rm PS}$ is broken to $G_{\rm SM}$ already
during inflation and thus no topological defects
can form at the end of inflation. Consequently,
the monopole problem is solved.

\par
It has been shown \cite{nshift} that shifted
hybrid inflation can be realized within the SUSY
PS model even without invoking any
non-renormalizable superpotential terms provided
that we supplement the model with some extra
Higgs superfields. This extension of the SUSY PS
model was actually introduced \cite{quasi} (see
also Ref.~\cite{quasitalks}) for a very different
reason. It is well known \cite{hall} that, in
SUSY models with exact Yukawa unification (or
with large $\tan\beta$ in general), such as the
simplest SUSY PS model, and universal boundary
conditions, the $b$-quark mass $m_b$
receives large SUSY corrections, which, for
$\mu>0$, lead to unacceptably large values of
$m_b$. Therefore, Yukawa unification must be
(moderately) violated so that, for $\mu>0$, the
predicted bottom quark mass resides within the
experimentally allowed range even with universal
boundary conditions. This requirement forces us
to extend the superfield content of this model
by including, among other superfields, an extra
pair of ${\rm SU}(4)_c$ non-singlet
${\rm SU}(2)_{\rm L}$ doublets, which naturally
develop \cite{wetterich} subdominant vacuum
expectation values (VEVs) and mix with the main
electroweak doublets of the model leading to a
moderate violation of Yukawa unification. (Note,
in passing, that this mechanism applied to the
$\mu<0$ case, where Yukawa unification predicts
a $m_b$ which after SUSY corrections becomes
unacceptably low, does not lead \cite{quasineg}
to a viable scheme.) It is remarkable that the
resulting extended SUSY PS model automatically
and naturally leads \cite{nshift} to a new
version of shifted hybrid inflation based solely
on renormalizable superpotential terms. This
inflationary scenario was called new shifted
hybrid inflation.

\par
In this paper, we show that the same extension of
the SUSY PS model can lead to a new version of
smooth hybrid inflation based only on
renormalizable superpotential terms provided that
a particular parameter of its superpotential is
adequately small. Indeed, the scalar potential of
the model, for a wide range of its other
parameters, possesses a valley of minima which
has an inclination already at the classical
level and can be used as inflationary path
leading to a novel realization of smooth hybrid
inflation.  This scenario will be referred to as
new smooth hybrid inflation. The predictions of
this inflationary model can be easily made
compatible with the recent three-year Wilkinson
microwave anisotropy probe (WMAP) measurements
\cite{wmap3} for natural values of the parameters
of the model. In particular, in global SUSY, the
spectral index turns out to be adequately small
so that it is consistent with the fitting of the
WMAP data \cite{wmap3} by the standard power-law
cosmological model with cold dark matter and a
cosmological constant ($\Lambda$CDM). Finally, as
in the ``conventional'' realization of smooth
hybrid inflation, $G_{\rm PS}$ is already broken
to $G_{\rm SM}$ during new smooth hybrid
inflation and, thus, no topological defects are
formed at the end of inflation.

\par
The inclusion of supergravity (SUGRA) corrections
with minimal K\"{a}hler potential raises the
spectral index above the allowed range as in
standard and shifted hybrid inflation for
relatively large values of the relevant
dimensionless coupling constant and in smooth
hybrid inflation for GUT breaking scale close to
its SUSY value (see Ref.~\cite{senoguz}).
However, the introduction of a non-minimal term
in the K\"{a}hler potential with appropriately
chosen sign can help to reduce the spectral index
so that it becomes comfortably compatible with
the data (compare with
Refs.~\cite{osamu,Bastero-Gil:2006cm,urRehman:2006hu}).
This can be achieved with the potential remaining
a monotonically increasing function of
the inflaton field everywhere on the inflationary
path. So, complications
\cite{Bastero-Gil:2006cm,urRehman:2006hu} from
the appearance of a local maximum and minimum of
the potential on the inflationary path when such
a non-minimal K\"{a}hler potential is used are
avoided. One possible complication is that the
system gets trapped near the minimum of the
inflationary potential and, consequently, no
hybrid inflation takes place. Another
complication is that, even if hybrid inflation of
the so-called hilltop type
\cite{lofti} occurs with the inflaton rolling
from the region of the maximum down to smaller
values, the spectral index can become compatible
with the data only at the cost of a mild tuning
of the initial conditions (see Ref.~\cite{gpp}).

\par
The paper is organized as follows. In
Sec.~\ref{sec:newsmooth}, we briefly introduce
the extended SUSY PS model and show that it
possesses a valley of minima along which
successful new smooth hybrid inflation can take
place. In Sec.~\ref{sec:sugra}, we discuss how
our new smooth inflationary scenario is affected
by the SUGRA corrections to the scalar potential.
Finally, in Sec.~\ref{sec:conclusions}, we
summarize our conclusions.

\section{New smooth hybrid inflation in global
supersymmetry}
\label{sec:newsmooth}

\par
We consider the extended SUSY PS model of
Ref.~\cite{quasi} as our starting point. This
model allows a moderate violation of
the asymptotic Yukawa unification so that, for
$\mu>0$, an acceptable value of the $b$-quark
mass is obtained even with universal boundary
conditions. The breaking of $G_{\rm PS}$ to
$G_{\rm SM}$ is achieved by the superheavy VEVs
($=M_{\rm GUT}
\simeq 2.86\ten{16}\units{GeV}$, the SUSY
GUT scale) of the right handed neutrino type
components of a conjugate pair of Higgs
superfields $H^c$ and $\bar{H}^c$ belonging to
the $(\bar{4},1,2)$ and $(4,1,2)$ representations
of $G_{\rm PS}$ respectively. The model also
contains a gauge singlet $S$ and a conjugate pair
of superfields $\phi$, $\bar{\phi}$ belonging to
the (15,1,3) representation of $G_{\rm PS}$. The
superfield $\phi$ acquires a (subdominant) VEV
which breaks $G_{\rm PS}$ to $G_{\rm SM}\times
{\rm U}(1)_{B-L}$. For details on the full field
content and superpotential, the global
symmetries, the charge assignments, and the
phenomenological and cosmological properties of
this model, the reader is referred to
Refs.~\cite{quasi,shift} (see also
Ref.~\cite{quasitalks}).

\par
As already mentioned, this extended SUSY PS model
leads \cite{nshift} to a new version of shifted
hybrid inflation which is based solely on
renormalizable interactions. The superpotential
terms which are relevant for this inflationary
scenario have been given in Eq.~(2.1) of
Ref.~\cite{nshift} and have been used there with
a particular choice of the phases of their
coupling constants. These terms with a different
(more convenient for our purposes here) choice of
basic parameters and their phases can be written
as
\beq\label{eq:superpotential}
W=\ka S(M^2-\phi^2)-\gamma S H^c\Hb^c+m\phi\pb
-\la\pb H^c\Hb^c,
\eeq
where $M$, $m>0$ are superheavy masses of the
order of $M_{\rm GUT}$ and $\ka$, $\gamma$,
$\la>0$ are dimensionless coupling constants.
These parameters are normalized so that they
correspond to the couplings between the SM
singlet components of the superfields. In a
general superpotential of the type in
Eq.~(\ref{eq:superpotential}), $M$, $m$ and any
two of the three dimensionless parameters $\ka$,
$\gamma$, $\la$ can always be made real and
positive by appropriately redefining the
phases of the superfields. The third
dimensionless parameter, however, remains
generally complex. For definiteness, we have
chosen here this parameter to be real and
positive too. One can show that the
superpotential in Eq.~(2.1) of Ref.~\cite{nshift}
with the particular choice of the phases of its
parameters considered there can become equivalent
to the superpotential in
Eq.~(\ref{eq:superpotential}) provided that its
real and positive parameter $\lambda$ is rotated
to the negative imaginary axis. Actually, the
form of the superpotential in
Eq.~(\ref{eq:superpotential}) can be derived
from the one in Eq.~(2.1) of Ref.~\cite{nshift}
by the replacement: $S\to -S$, $\phi\to i\phi$,
$\pb\to -i\pb$, $\ka\to \gamma$, $\beta\to \ka$,
$\la\to -i\la$, $M^2\to (\ka/\gamma)M^2$.

\par
In this paper, we will show that the specific
superpotential of Eq.~\eqref{eq:superpotential}
leads to a new version of smooth hybrid inflation
\cite{smooth} provided that the parameter
$\gamma$ is taken to be adequately small. To this
end, we will first examine the case with $\gamma$
set to zero and then we will move on to allow a
small, but non-zero value for this parameter. Note
that one could get rid of the $\gamma$-term in
the superpotential entirely by introducing an
extra $Z_2$ symmetry under which $H^c$, $\phi$,
and $\pb$ change sign. However, this would
disallow the solution of the $b$-quark mass
problem \cite{quasi} and, thus, invalidate the
original motivation for introducing this extended
SUSY PS model. This is due to the fact that the
superpotential term which generates the crucial
mixing between the ${\rm SU}(4)_c$ singlet and
non-singlet ${\rm SU}(2)_{\rm L}$ doublets (see
Ref.~\cite{quasi}) is forbidden by this discrete
symmetry. Needless to say that, for $\gamma=0$,
all the choices for the phases of the parameters
in Eq.~(\ref{eq:superpotential}) are equivalent.

\subsection{The $\gamma=0$ case}
\label{sec:g0}

\par
Setting $\gamma=0$, the F-term scalar potential
obtained from $W$ is given by
\bea
V&=&\ka^2|M^2-\phi^2|^2+|m\pb-2\ka S\phi|^2
\nonumber\\
& &+|m\phi-\la H^c\Hb^c|^2+\la^2|\pb|^2
\left(|H^c|^2+|\Hb^c|^2\right),
\label{eq:Fpot}
\eea
where the complex scalar fields which belong to
the SM singlet components of the superfields are
denoted by the same symbol. We will ignore
throughout the soft SUSY breaking terms
\cite{sstad} in the scalar potential since their
effect on inflationary dynamics is negligible in
our case as in the case of the conventional
realization of smooth hybrid inflation (see
Ref.~\cite{urRehman:2006hu}).

\par
From the potential in Eq.~(\ref{eq:Fpot}), we
find that the SUSY vacua lie at
\beq
\pb=S=0,\quad \phi^2=M^2 ,\quad
H^c\Hb^c=\frac{m}{\la}\,\phi.
\eeq
The vanishing of the D-terms yields
$\Hb^{c*}=e^{i\theta}H^c$, which implies that we
have four distinct SUSY vacua:
\bea
\phi=M,\quad H^c=\Hb^c=\pm\sqrt{\frac{mM}{\la}}
\quad (\theta=0),\label{vacua+}\\
\phi=-M,\quad H^c=-\Hb^c=\pm\sqrt{\frac{mM}{\la}}
\quad (\theta=\pi) \label{vacua-}
\eea
with $\pb=S=0$. Here, for simplicity, $H^c$,
$\Hb^c$ have been rotated to the real axis by an
appropriate gauge transformation. However, we
should keep in mind that the fields $H^c$,
$\pm\Hb^{c*}$ (the plus or minus sign corresponds
to $\theta=0$ or $\pi$ respectively) can have an
arbitrary common phase in the vacuum and, thus,
the two distinct vacua in Eq.~\eqref{vacua+} or
\eqref{vacua-} are not, in reality, discrete,
but rather belong to a continuous $S^1$ vacuum
submanifold. Note that the vacua in
Eq.~(\ref{vacua+}) are related to the ones in
Eq.~(\ref{vacua-}) by the $Z_2$ symmetry
mentioned above. As we will see later, the
specific point of the vacuum manifold towards
which the system is heading is already chosen
during inflation. So the model does not encounter
any topological defect problem. Actually, there
is no production of topological defects at all.

\par
It is not very hard to show that, at any possible
minimum of the potential, $\eps=0$ or $\pi$ and
$\eps=\bar{\eps}=-\theta$, where $\eps$ and
$\bar{\eps}$ are the phases of $\phi$ and $\pb$
respectively ($S$ can be made real by an
appropriate global ${\rm U}(1)$ R
transformation). This
remains true even at the minima of $V$ with
respect to $\phi$, $\pb$, $H^c$, and $\Hb^c$ for
fixed $S$. So, we will restrict ourselves to
these values of $\theta$ and the phases of $\phi$
and $\pb$. The scalar potential then takes the
form
\bea\label{eq:Vmin}
V_{\rm min}
&=&\ka^2\left(|\phi|^2-M^2\right)^2+\left(2\ka
|S||\phi|-m|\pb|\right)^2 \nonumber\\
& &+\left(m|\phi|-\la |H^c|^2\right)^2+2\la^2
|\pb|^2|H^c|^2.
\eea
The derivatives of this potential with respect to
the norms of the fields are
\bea
\frac{\partial V_{\rm min}}{\partial |S|}
&=& 4\ka\left(2\ka|S||\phi|-m|\pb|\right)|\phi|,
\label{eq:dVdS}\\
\frac{\partial V_{\rm min}}{\partial |\phi|}
&=& 4\ka^2\left(|\phi|^2-M^2\right)|\phi|
+4\ka\left(2\ka |S||\phi|-m|\pb|\right)|S|
\nonumber\\
& &+2m\left(m|\phi|-\la|H^c|^2\right),
\label{eq:dVdphi}\\
\frac{\partial V_{\rm min}}{\partial |\pb|}
&=& -2m\left(2\ka|S||\phi|-m|\pb|\right)+4\la^2
|\pb||H^c|^2,
\label{eq:dVdpb}\\
\frac{\partial V_{\rm min}}{\partial |H^c|}
&=& -4\la\left(m|\phi|-\la|H^c|^2-\la|\pb|^2
\right)|H^c|.
\label{eq:dVdHc}
\eea

\par
The potential $V_{\rm min}$ possesses two flat
directions. The first one is the trivial flat
direction at $|\phi|=|\pb|=|H^c|=0$ with
$V=V_{\rm tr}\equiv\ka^2M^4$. The second one
exists only if $\tilde{\mu}^2\equiv M^2-m^2/2
\ka^2>0$ and is a shifted flat direction at
\beq
|\phi|=\tilde{\mu},\quad
|\pb|=\frac{2\ka\tilde{\mu}}{m}\,|S|,\quad
|H^c|=0,
\eeq
where $\tilde{\mu}\equiv (M^2-m^2/2\ka^2)^{1/2}$,
with \mbox{$V=\ka^2(M^4-\tilde{\mu}^4)$}. The
mass-squared matrix of the variables $|S|$,
$|\phi|$, $|\pb|$, and $|H^c|$ on the trivial
flat direction is
\beq \label{eq:mass2}
\left(\ba{cccc}
0 & 0 & 0 & 0 \\
0 & 4\ka^2(2|S|^2-\tilde{\mu}^2) & -4\ka m|S| & 0
\\
0 & -4\ka m|S| & 2m^2 & 0 \\
0 & 0 & 0 & 0
\ea\right).
\eeq
If $M_{\phi\pb}$ denotes the $|\phi|$, $|\pb|$
sector of this matrix, then
\bea
\det(M_{\phi\pb}) &=& -8\ka^2m^2\tilde{\mu}^2, \\
\tr(M_{\phi\pb}) &=& 4\ka^2(2|S|^2-\tilde{\mu}^2)
+2m^2.
\eea
So, the matrix $M_{\phi\pb}$ has one positive and
one negative eigenvalue for $\tilde{\mu}^2>0$ and
two positive eigenvalues for $\tilde{\mu}^2<0$.
In the former case, the trivial flat direction is
a path of saddle points and the shifted flat
direction is an honest candidate for the
inflationary path. However, in this paper, we
will concentrate on the latter case and set
$\mu^2\equiv -\tilde{\mu}^2>0$. Note that, even
in this case, the trivial flat direction may not
be a valley of local minima because of the
existence of the zero eigenvalue of the full
mass-squared matrix in Eq.~\eqref{eq:mass2}
associated with the field $|H^c|$. It is
perfectly conceivable that, starting from any
point on the trivial flat direction, there exist
paths along which the potential decreases as we
move away from this flat direction (at least
initially). Actually, as we will show below, this
happens to be the case here.

\par
To examine the stability of the trivial flat
direction, we consider a point on it and try to
see whether, starting from this point, one can
construct paths in the
\mbox{$\left(|H^c|,|\phi|,|\pb|\right)$} space
along which the potential in Eq.~(\ref{eq:Vmin})
has a local maximum at the point on the trivial
flat direction. In particular, we will try to
find the path of steepest descent. Throughout the
analysis, $|S|$ will be considered as a fixed
parameter characterizing the chosen point on the
trivial flat direction rather than as a dynamical
variable. Setting $|H^c|=\chi$, $|\phi|=\psi$,
and $|\pb|=\omega$, we can parameterize any
path in the field space as
\mbox{$\left(\chi,\psi(\chi),\omega(\chi)
\right)$}. We see, from
the form of the matrix in Eq.~\eqref{eq:mass2},
that the required paths must be tangential to
the $|H^c|$ direction at their origin (because,
for $\mu^2>0$, displacement along the $|\phi|$ or
$|\pb|$ direction enhances the potential
locally). Thus, the required initial conditions
for these paths are
\beq \label{eq:initcond1}
\chi=0,\quad \psi(0)=\omega(0)=0,\quad
\psi'(0)=\omega'(0)=0,
\eeq
where prime denotes derivation with respect to
$\chi$.

\par
The potential $V_{\rm min}$ on such a path can
be written as
\beq
F(\chi)=f\left(\chi,\psi(\chi),\omega(\chi)
\right),
\eeq
where $f(\chi,\psi,\omega)\equiv
V_{\rm min}(\chi,\psi,\omega)$. It is then
obvious that $F'(0)$ is zero by construction
since
\beq\label{eq:initcond2}
\left(\bar\nabla V_{\rm min}\right)_0=0,
\eeq
where $(...)_0$ denotes the value at
$\chi=\psi=\omega=0$.
Thus, the initial point of the path is a critical
point of $F(\chi)$ (as it should). Moreover, it
is easily verified, using Eqs.~\eqref{eq:mass2},
\eqref{eq:initcond1}, and \eqref{eq:initcond2},
that $F''(0)=0$, which means that we cannot
decide on the stability of the trivial flat
direction merely from the mass-squared matrix
in Eq.~\eqref{eq:mass2}. Therefore, higher
derivatives of $F(\chi)$ must be considered. We
find that $F'''(0)=0$ and
\beq\label{eq:F4}
F''''(0)=\alpha+\zeta\psi''_0+\rho\omega''_0+
(\psi''_0,\omega''_0)
\left(\ba{cc}
a & c \\
c & b
\ea\right)
\left(\ba{c}\psi''_0\\\omega''_0\ea\right)
\eeq
with $\psi''_0\equiv\psi''(0)$,
$\omega''_0\equiv\omega''(0)$,
\bea
\alpha &\equiv&\left(\frac{\partial^4 f}
{\partial \chi^4}\right)_0=24\la^2, \nonumber\\
\zeta &\equiv&6\left(\frac{\partial^3 f}{\partial
\chi^2\partial \psi}\right)_0=-24\la m,
\nonumber\\
\rho &\equiv& 6\left(\frac{\partial^3 f}{\partial
\chi^2\partial \omega}\right)_0=0, \nonumber\\
a &\equiv& 3\left(\frac{\partial^2 f}{\partial \psi^2}
\right)_0=12\ka^2(\mu^2+2|S|^2), \nonumber\\
b &\equiv&3\left(\frac{\partial^2 f}{\partial\omega^2}
\right)_0=6m^2, \nonumber\\
c &\equiv& 3\left(\frac{\partial^2 f}{\partial\psi
\partial\omega}\right)_0=-12\ka m|S|, \nonumber
\eea
where Eqs.~(\ref{eq:Vmin}), \eqref{eq:initcond1},
and \eqref{eq:initcond2} were used. Note that the
$2\times 2$ matrix in the last term in the right
hand side of Eq.~\eqref{eq:F4} is just
$3M_{\phi\pb}$, which is positive definite for
$\mu^2>0$ (see the discussion following
Eq.~\eqref{eq:mass2}).

\par
By applying the transformation
\beq\label{eq:trans1}
\psi''_0=\hat{\psi}''_0+\delta\psi''_0,\quad
\omega''_0=\hat{\omega}''_0+\delta \omega''_0,
\eeq
one can show that Eq.~\eqref{eq:F4} can be
brought into the form
\beq\label{eq:F4trans}
F''''(0)=-\frac{24\la^2M^2}{\mu^2}+
(\delta\psi''_0,\delta \omega''_0)
\left(\ba{cc}
a & c \\
c & b
\ea\right)
\left(\ba{c}\delta\psi''_0 \\
\delta\omega''_0\ea\right)
\eeq
with
\beq
\hat{\psi}''_0=-\frac{\zeta b}{2(ab-c^2)}>0,
\quad
\hat{\omega}''_0=\frac{\zeta c}{2(ab-c^2)}
\geq 0.
\eeq
The last term in the right hand side of
Eq.~\eqref{eq:F4trans} is a
positive definite quadratic form in
\mbox{$\delta\psi''_0\geq -\hat{\psi}''_0$},
\mbox{$\delta\omega''_0\geq -\hat{\omega}''_0$}
(the non-positive lower bounds originate from the
fact that $\psi''_0$, $\omega''_0\geq 0$, which
in turn comes from Eq.~\eqref{eq:initcond1} and
the fact that $\psi$, $\omega\geq 0$ by their
definition). It is obvious then that there exist
choices of $\delta\psi''_0$, $\delta\omega''_0$
which render $F''''_0$ negative. Thus, on the
corresponding paths, $F(\chi)$ has a local
maximum at $\chi=0$. We conclude that the trivial flat
direction is a path of saddle points rather
than a valley of local minima. The path of
steepest decent corresponds to $\delta\psi''_0$,
$\delta\omega''_0=0$, which minimizes $F''''_0$.

\par
We have just seen that, for any fixed value of
$|S|$, $V_{\rm min}$ has a local maximum on the
trivial flat direction at $|\phi|=|\pb|=|H^c|=0$.
Moreover, $V_{\rm min}\to +\infty$ as
$|\phi|^2+|\pb|^2+|H^c|^2\to\infty$. This
means that, for each value of $|S|$,
$V_{\rm min}$ must have a non-trivial absolute
minimum (where at least one the fields $|\phi|$,
$|\pb|$, and $|H^c|$ has a non-zero value). These
minima then form a valley, which may be used as
inflationary trajectory. Actually, as we will
show soon, this trajectory is not flat and
resembles the path used in Ref.~\cite{smooth}
for smooth hybrid inflation. We can find the
valley of minima of $V_{\rm min}$ by minimizing
this potential with respect to $|\phi|$, $|\pb|$,
and $|H^c|$, regarding $|S|$ as a fixed
parameter. This amounts to solving the system of
equations that is formed by equating the partial
derivatives in
Eqs.~\eqref{eq:dVdphi}-\eqref{eq:dVdHc} with
zero. We obtain three non-linear equations with
three unknowns, which cannot be solved
analytically. Though, as in the case of
conventional smooth hybrid inflation
\cite{smooth}, we will try to find a solution in
the large $|S|$ limit. In particular, we will
try to find a power series solution with respect
to some parameter of the form
$\text{``mass''}/|S|$ which remains smaller than
unity throughout the entire range of $|S|$ which
is relevant for inflation. As it will become
clear below, a convenient quantity for the
``mass'' in the numerator is
$v_g\equiv\sqrt{mM/\lambda}$, which is just the
VEV $|\vev{H^c}|$ at the SUSY minima of the
potential. Re-expressing the system of equations
by using the dimensionless variables
$x\equiv|\phi|/M$, $y\equiv|\pb|/\sqrt{2}p\,v_g$,
$z\equiv|H^c|/v_g$, and $w\equiv v_g/|S|$, where
$p\equiv\sqrt{2}\ka M/m$ is a dimensionless
parameter smaller than unity for $\mu^2>0$, we
obtain
\begin{gather}
wx(x^2-1)+4yz^2+2wy^2=0, \nonumber\\
x-wy=\sqrt{2}\,\frac{\la}{\ka}\,pwyz^2,
\label{eq:system}\\
x=z^2+2p^2y^2. \nonumber
\end{gather}
Writing the variables $x$, $y$, and $z^2$ as power
series in $w$ and equating the coefficients of
the corresponding powers of $w$ in the two sides
of Eqs.~\eqref{eq:system}, we get
\bea
x &=& x_2w^2+x_4w^4+\dots, \nonumber\\
y &=& y_1w+y_3w^3+\dots,
\label{eq:VarExpansions}\\
z^2 &=& z_2w^2+z_4w^4+\dots, \nonumber
 \eea
where the coefficients $x_i$, $y_i$, and $z_i$
depend only on the parameter $p$ and the ratio
$\la/\ka$ and are given by
\begin{gather}
x_2=y_1=\frac{3}{8p^2}\left(1-\sqrt{1-8p^2/9}
\right),\label{eq:xyz1}\\
z_2=\frac{1}{4}\,(1-2x_2),\label{eq:xyz2}\\
x_4=\frac{\sqrt{2}}{8}\,\frac{\la}{\ka}\,p\;
\frac{x_2(1-2x_2)(3-10x_2)}{1-3x_2},\label{eq:xyz3}\\
y_3=\frac{1-4x_2}{3-10x_2}\;x_4,\label{eq:xyz4}\\
z_4=\frac{1+2(1-2p^2)x_2}{3-10x_2}\;x_4.
\label{eq:xyz}
\end{gather}

\par
A useful approximation to these coefficients can
be found by expanding them with respect to the
small parameter $p$ (see below). Thus, to first
non-trivial order in $p$, we find the following
simple expressions:
\begin{gather}
x_2=y_1=z_2=\frac{1}{6}, \label{eq:p1}\\
x_4=z_4=\frac{\sqrt{2}}{27}\,\frac{\la}{\ka}\,p,
\quad y_3=\frac{\sqrt{2}}{108}\,\frac{\la}{\ka}\,p
\label{eq:p2}
\end{gather}
and Eq.~\eqref{eq:VarExpansions} takes the form
\bea
|\phi| &\simeq& \frac{Mv_g^2}{6|S|^2}
\left(1+\frac{2\sqrt{2}}{9}\,\frac{\la}{\ka}\,p\,
w^2+\dots\right),\nonumber\\
|\pb| &\simeq& \sqrt{2}p\,\frac{v_g^2}{6|S|}
\left(1+\frac{\sqrt{2}}{18}\,\frac{\la}{\ka}\,p\,
w^2+\dots\right), \label{eq:FieldExpansions}\\
|H^c| &\simeq& \frac{v_g^2}{\sqrt{6}\,|S|}
\left(1+\frac{\sqrt{2}}{9}\,\frac{\la}{\ka}\,p\,
w^2+\dots\right). \nonumber
\eea
Taking into account the possible values of the
phases $\eps$, $\bar{\eps}$, and $\theta$ (and
with $H^c$, $\bar{H}^c$ rotated to the real
axis), we see that the potential in
Eq.~\eqref{eq:Fpot} possesses four valleys of
absolute minima (for fixed $|S|$) which
presumably lead to the four SUSY vacua in
Eqs.~\eqref{vacua+} and \eqref{vacua-}. We
should keep in mind, though, that the two
valleys corresponding to the same value of
$\theta$ are not discrete, but continuously
connected since $H^c$, $\pm\bar{H}^{c*}$ can have
an arbitrary common phase. The expansions in
Eq.~\eqref{eq:FieldExpansions} hold as long as
$w<1$, that is $|S|>v_g$. In the following, we
will keep only the terms of leading order in
$w$ in the above equations. Although this might
seem somewhat arbitrary, we will justify it
later. Substituting the expansions in
Eq.~\eqref{eq:FieldExpansions} into the potential
of Eq.~\eqref{eq:Vmin} and keeping only terms of
leading order in $w$, we get
\beq\label{eq:Veff}
V_{\text{min}}\simeq
\ka^2M^4\left(1-\frac{v_g^4}{54|S|^4}\right).
\eeq
This is exactly the form of the potential for
smooth hybrid inflation considered in
Ref.~\cite{smooth}. Thus, we have shown that
the present model possesses inflationary paths
leading to smooth hybrid inflation. We will
call the resulting scenario new smooth hybrid
inflation since, in contrast to the
conventional realization of smooth hybrid
inflation, it is achieved by using only
renormalizable interactions. It is evident that,
as the system follows the new smooth
inflationary path, the phases of the various
fields remain fixed. Moreover, the particular
point of the vacuum manifold towards which the
system is heading is already chosen during
inflation and we encounter no cosmological
defect problems.

\par
Setting $S=\si/\sqrt{2}$, where $\si$ is the
canonically normalized real inflaton field
(recall that $S$ was made real by an R
transformation), we obtain the potential along
the new smooth inflationary path
\beq
V\simeq \label{eq:Vg0}
v_0^4\left(1-\frac{2v_g^4}{27\si^4}\right),
\eeq
where $v_0\equiv\sqrt{\ka}M$ is the inflationary
scale. The slow-roll parameters $\veps$, $\eta$
and the parameter $\xi^2$, which enters the
running of the spectral index, are (see e.g.
Ref.~\cite{review})
\bea
\veps &\equiv& \frac{m_{\rm P}^2}{2}\,
\left(\frac{V^{(1)}(\si)}{V(\si)}\right)^2
\simeq\frac{32m_{\rm P}^2v_g^8}{729\si^{10}},\\
\eta &\equiv& m_{\rm P}^2\,
\left(\frac{V^{(2)}(\si)}{V(\si)}\right)
\simeq-\frac{40m_{\rm P}^2v_g^4}{27\si^6},\\
\xi^2 &\equiv& m_{\rm P}^4
\left(\frac{V^{(1)}(\si)V^{(3)}(\si)}{V^2(\si)}
\right)
\simeq\frac{640m_{\rm P}^4v_g^8}{243\si^{12}},
\eea
where the superscript $(n)$ denotes the $n$-th
derivative with respect to $\si$ and $m_{\rm P}$
is the reduced Planck mass. Inflation ends at
$\si=\si_f$ (taken positive by an R
transformation) where $\eta=-1$, which gives
\beq
\si_f^6\simeq\frac{40m_{\rm P}^2v_g^4}{27}.
\label{eq:sigmaf}
\eeq

\par
The number of e-foldings from the time when the
pivot scale $k_0=0.002\units{Mpc}^{-1}$ crosses
outside the inflationary horizon until the end of
inflation is given (see e.g. Ref.~\cite{review})
by
\beq
N_Q\simeq\frac{1}{m_{\rm P}^2}\,
\int_{\si_f}^{\si_Q}
\frac{V(\si)}{V^{(1)}(\si)}\,d\si
\simeq\frac{9}{16m_{\rm P}^2v_g^4}\,
\left(\si_Q^6-\si_f^6\right),
\eeq
where $\si_Q\equiv\sqrt{2} S_Q>0$ is the value
of the inflaton field at horizon crossing of the
pivot scale. Taking into account the fact that
$\si_f\ll\si_Q$, we can write
\beq
\si_Q^6\simeq\frac{16m_{\rm P}^2v_g^4}{9}\,N_Q.
\label{eq:sigmaQ}
\eeq
The power spectrum $P_{\mathcal R}$ of the
primordial curvature perturbation at the scale
$k_0$ is given (see e.g. Ref.~\cite{review}) by
\beq\label{eq:Perturbations}
P_{\mathcal R}^{1/2}\simeq
\frac{1}{2\pi\sqrt{3}}\,\frac{V^{3/2}(\si_Q)}
{m_{\rm P}^3V^{(1)}(\si_Q)}\simeq
\frac{3^{5/6}N_Q^{5/6}}{2^{2/3}\pi}\,
\left(\frac{v_0^3}{m_{\rm P}^2v_g}\right)^{2/3}.
\eeq
The spectral index $n_s$, the tensor-to-scalar
ratio $r$, and the running of the spectral index
$dn_s/d\ln k$ are given (see e.g.
Ref.~\cite{review}) by
\begin{gather}
n_s\simeq 1+2\eta-6\veps\simeq1-\frac{5}{3N_Q},
\nonumber\\
r\simeq\,16\veps\simeq\frac{2^{7/3}}{3^{8/3}
N_Q^{5/3}}\,\left(\frac{v_g}{m_{\rm P}}
\right)^{4/3},\\
\frac{dn_s}{d\ln k}\simeq16\veps\eta-24\veps^2
-2\xi^2\simeq -\frac{5}{3N_Q^2},\nonumber
\end{gather}
where $\veps$, $\eta$, and $\xi^2$ are evaluated
at $\si=\si_Q$. The number of e-foldings $N_Q$
required for solving the horizon and flatness
problems of standard hot big bang cosmology is
given (see e.g. Ref.~\cite{lectures})
approximately by
\beq\label{eq:NQvsVinf}
N_Q\simeq53.76\,+\frac{2}{3}\,\ln\left(\frac{v_0}
{10^{15}\units{GeV}}\right)+\frac{1}{3}\,\ln
\left(\frac{T_r}{10^9\units{GeV}}\right),
\eeq
where $T_r$ is the reheat temperature which is
expected not to exceed about $10^9\units{GeV}$,
which is the well-known gravitino bound
\cite{gravitino}.

\par
Taking $v_g$ to have the SUSY GUT value, i.e.
$v_g\simeq 2.86\ten{16}\units{GeV}$ (see below),
$T_r$ to saturate the gravitino bound, i.e. $T_r\simeq
10^9\units{GeV}$, and the WMAP \cite{wmap3}
normalization $P_{\mathcal R}^{1/2}\simeq 4.85
\ten{-5}$ at the comoving scale $k_0$, we can
solve Eqs.~\eqref{eq:Perturbations} and
\eqref{eq:NQvsVinf} numerically. We obtain
\beq
N_Q\simeq53.78,\quad v_0
\simeq1.036\ten{15}\units{GeV}.
\eeq
The spectral index, the tensor-to-scalar ratio,
and the running of the spectral index are then
\beq
n_s\simeq 0.969,\;\;
r\simeq 9.4\ten{-7},\;\;
\frac{dn_s}{d\ln k}\simeq -5.8\ten{-4}.
\label{eq:nsralpha}
\eeq
We see that the running of the spectral index
and the tensor-to-scalar ratio are negligible
and, thus, the standard power-law $\Lambda$CDM
cosmological model should hold to a very
good accuracy. Fitting the three-year results
from WMAP \cite{wmap3} with this cosmological
model, one obtains that, at the pivot scale
$k_0$,
\begin{equation}\label{eq:nswmap}
n_{\rm s}=0.958\pm 0.016~\Rightarrow~0.926
\lesssim n_{\rm s} \lesssim 0.99
\end{equation}
at 95$\%$ confidence level. So, the value of the
spectral index in Eq.~\eqref{eq:nsralpha} is
perfectly acceptable. It is, actually, the same
as in conventional smooth hybrid inflation
\cite{smooth} since the inflationary potential
for large $|S|$ is exactly the same, as we
already pointed out.

\par
We have already fixed the values of the
parameters $v_0=\sqrt{\kappa}M$ and
$v_g=\sqrt{mM/\lambda}$. So, we are free to
make two more choices in order to determine the
four parameters of the model $m$, $M$, $\ka$,
and $\la$. A legitimate choice is to set
$\ka=\la$ and $m=M$ which leads to quite
natural values for the parameters, namely
\begin{gather}
m=M=\sqrt{v_0v_g}\simeq5.44\ten{15}\units{GeV},
\nonumber\\
\ka=\la=\frac{v_0}{v_g}\simeq0.0362.
\label{eq:parameters}
\end{gather}
For these values, we find, from
Eqs.~\eqref{eq:sigmaf} and \eqref{eq:sigmaQ}, that
$\sigma_f\simeq 1.34\ten{17}\units{GeV}$ and
$\sigma_Q\simeq 2.69\ten{17}\units{GeV}$.

\par
\begin{figure}[tp]
\centering
\includegraphics[width=\linewidth]{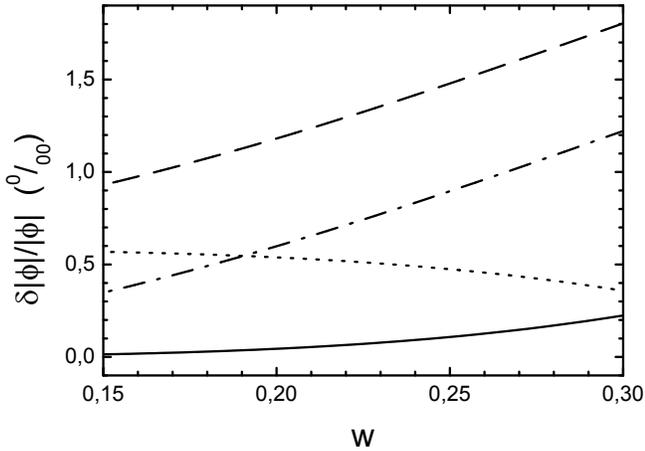}
\caption{Relative error in $|\phi|$ on the new
smooth inflationary path in global SUSY for the
values of the parameters in
Eq.~\eqref{eq:parameters} and $\gamma=0$ when we
use the expansion of Eq.~\eqref{eq:VarExpansions}
up to second order in $w$ with coefficient
evaluated to leading order in $p$ (dashed line) or
accurately (dot-dashed line) and up to fourth
order in $w$ with coefficients evaluated to
leading order in $p$ (dotted line) or accurately
(solid line).}
\label{fig:RelativeError}
\end{figure}

Let us now turn to the justification of the
expansions in Eqs.~\eqref{eq:VarExpansions} and
\eqref{eq:FieldExpansions}. The value of $|S|$
at the termination of inflation is approximately
given by
\beq
S_f^6=\frac{\si_f^6}{2^3}\simeq
\frac{5m_{\rm P}^2v_g^4}{27}.
\eeq
Therefore, the maximum value of $w$ during
inflation is
\beq
w_{\max}=\frac{v_g}{S_f}\simeq\frac{3^{1/2}}
{5^{1/6}}\left(\frac{v_g}{m_{\rm P}}\right)^{1/3}
\simeq0.3.
 \eeq
Consequently, the condition $w<1$ is well
satisfied during inflation and the expansions in
Eq.~\eqref{eq:VarExpansions} are valid. Moreover,
$p\simeq 0.0512\ll 1$ for the values in
Eq.~\eqref{eq:parameters} and, thus, for
$\la\sim\ka$, the expansions in
Eq.~\eqref{eq:FieldExpansions} are also
justified. We find numerically that these
expansions are, actually, justified in the entire
range $w\leq w_{\max}$ even for values of $p$
close to unity and $\la>\ka$. Rough estimates of
the maximum relative errors when only the leading
order term is kept in the expansions of
Eq.~\eqref{eq:FieldExpansions} are given by the
second term in the parentheses in this equation
for $w=w_{\max}$. For the values in
Eq.~\eqref{eq:parameters}, we get that the
maximum relative error in $|\phi|$, which seems
to be the largest of the errors in $|\phi|$,
$|\pb|$, and $|H^c|$, is given by the estimate
\beq\label{eq:ErrorInPhi}
\frac{\delta|\phi|}{|\phi|}
\simeq \frac{2\sqrt{2}}{9}\,\frac{\la}{\ka}\,p\;
w_{\max}^2\simeq 1.45\ten{-3}\sim
1\text{\textperthousand}.
\eeq
This is verified numerically as shown in
Fig.~\ref{fig:RelativeError}, where we plot the
relative error in $|\phi|$ during inflation when
we approximate the new smooth inflationary path
by the expansions in Eq.~\eqref{eq:VarExpansions}.
Note that, in order to retain a precision better
than $1\%$ in $|\phi|$ keeping only the leading
order term in its expansion in
Eq.~\eqref{eq:FieldExpansions}, the relation
$M\lsim v_g/2$ has to hold, as can be seen from
Eq.~\eqref{eq:ErrorInPhi} for
$w_{\max}\simeq 0.3$.

\par
The identification of $v_g$, which is the VEV
$|\vev{H^c}|$ or $|\vev{\bar{H}^c}|$, with the
SUSY GUT scale $M_{\rm GUT}$ can be easily
justified. As already mentioned, the VEVs of
$H^c$, $\bar{H}^c$ break the PS gauge group to
$G_{\rm SM}$, whereas the VEV of the field $\phi$
breaks it only to $G_{\rm SM}\times
{\rm U}(1)_{B-L}$. So, the gauge boson $A^\perp$
corresponding to the linear combination of
${\rm U}(1)_{Y}$ and ${\rm U}(1)_{B-L}$ which is
perpendicular to ${\rm U}(1)_{Y}$ acquires its
mass squared
$m^2_{A^\perp}=(5/2)g^2|\vev{H^c}|^2$ solely from
the VEVs of $H^c$, $\bar{H}^c$ ($g$ is the SUSY
GUT gauge coupling constant). On the other hand,
the masses squared $m_A^2$ and $m_{W_{\rm R}}^2$
of the color triplet, anti-triplet ($A^\pm$) and
charged ${\rm SU}(2)_{\rm R}$ ($W^\pm_{\rm R}$)
gauge bosons get contributions from $\vev{\phi}$
too. Namely, $m_A^2=g^2(|\vev{H^c}|^2+(4/3)
|\vev{\phi}|^2)$ and $m_{W_{\rm R}}^2=g^2
(|\vev{H^c}|^2+2|\vev{\phi}|^2)$. For the values
in Eq.~\eqref{eq:parameters}, however,
\beq
\frac{|\vev{\phi}|^2}{|\vev{H^c}|^2}=
\frac{\lambda M}{m}\simeq 0.0362\ll 1,
\eeq
which implies that $m_A\approx m_{W_{\rm R}}
\approx gv_g$ within a few per cent. So, $v_g$ is
approximately equal to the practically common
mass of the SM non-singlet superheavy gauge
bosons divided by $g\approx 0.7$, which is, in
turn, equal to $M_{\rm GUT}\simeq 2.86\ten{16}
\units{GeV}$ (the SM singlet gauge boson
$A^\perp$ does not affect the renormalization
group equations).

\subsection{The $\gamma\neq 0$ case}
\label{sec:gn0}

\par
We will now turn to the case of a non-vanishing,
but small value of the parameter $\gamma$. The
scalar potential, in this case, takes the form
\bea \label{eq:Fpotg}
V&=&|\ka\,(M^2-\phi^2)-\gamma H^c\Hb^c|^2
\nonumber\\
& &+|m\pb-2\ka S\phi|^2+|m\phi-\la H^c\Hb^c|^2
\nonumber\\
& &+|\gamma S+\la\pb\,|^2\left(|H^c|^2+|\Hb^c|^2
\right)
\eea
and the SUSY vacua lie at
\begin{gather}
\phi=\frac{\gamma m}{2\ka\la}\left(
-1\pm\sqrt{1+\frac{4\ka^2\la^2M^2}
{\gamma^2m^2}}\,\right)\equiv\phi_{\pm},\\
\pb=S=0,\quad H^c\Hb^c=\frac{m}{\la}\,\phi.
\end{gather}
Again, the vanishing of the D-terms yields
$\Hb^{c*}=e^{i\theta}H^c$, which implies that we
have four distinct SUSY vacua:
\bea
\phi=\phi_{+},\quad H^c=\Hb^c=\pm
\sqrt{\frac{m\phi_{+}}{\la}} \quad (\theta=0),
\label{eq:vacuum1}\\
\phi=\phi_{-},\quad H^c=-\Hb^c=\pm
\sqrt{\frac{-m\phi_{-}}{\la}} \quad
(\theta=\pi)
\label{eq:vacuum2}
\eea
with $\pb=S=0$. Here $H^c$, $\Hb^{c}$ are rotated
to the real axis, but we should again keep in
mind that the two vacua in Eq.~\eqref{eq:vacuum1}
or \eqref{eq:vacuum2} belong, in reality, to a
continuum of vacua. One can show that the
potential now generally possesses three flat
directions. The first one is the usual trivial
flat direction at $\phi=\pb=H^c=\Hb^c=0$ with
$V=V_{\rm tr}=\ka^2M^4$. The second one
exists only if $\tilde{\mu}^2>0$ and lies at
\beq
\phi=\pm\,\tilde{\mu},\quad
\pb=\frac{2\ka\phi}{m}\,S,\quad
H^c=\Hb^c=0.
\label{semishift}
\eeq
It is a shifted flat direction with
\mbox{$V=\ka^2(M^4-\tilde{\mu}^4)$} along which
$G_{\rm PS}$ is broken to $G_{\rm SM}\times
{\rm U}(1)_{B-L}$. Note that the positions of the
trivial and shifted flat directions remain the
same as in the $\gamma=0$ case. The third flat
direction, which appears at
\begin{gather}
\phi=-\frac{\gamma m}{2\ka\la},\quad
\pb=-\frac{\gamma}{\la}\,S,\\
H^c\Hb^c=\frac{\ka\gamma(M^2-\phi^2)+\la m\phi}
{\gamma^2+\la^2},\\
V=V_\text{nsh}\equiv\frac{\ka^2\la^2}
{\gamma^2+\la^2}\left(M^2+\frac{\gamma^2m^2}
{4\ka^2\la^2}\right)^2,
\end{gather}
exists only for $\gamma\neq 0$ and is analogous
to the trajectory for the new shifted hybrid
inflation of Ref.~\cite{nshift}. Along this
direction, $G_{\rm PS}$ is broken to
$G_{\rm SM}$. In our subsequent discussion, we
will again concentrate on the case where $\mu^2=
-\tilde{\mu}^2>0$. It is interesting to note
that, in this case, we always have
$V_\text{nsh}>V_{\rm tr}$ and it is, thus, more
likely that the system will eventually settle
down on the trivial rather than the new shifted
flat direction (the shifted flat direction in
Eq.~(\ref{semishift}) does not exist in this
case).

\par
If we expand the complex scalar fields $\phi$,
$\pb$, $H^c$, $\Hb^c$ in real and imaginary parts
according to the prescription $s=(s_1+i\,
s_2)/\sqrt{2}$, we find that, on the trivial flat
direction, the mass-squared matrices
$M_{\phi1}^2$ of $\phi_1$, $\pb_1$ and
$M_{\phi2}^2$ of $\phi_2$, $\pb_2$ are
\beq
M_{\phi1(\phi2)}^2=\left(\ba{cc}
m^2+4\ka^2|S|^2\mp2\ka^2M^2 & -2\ka m S \\
-2\ka m S  & m^2 \ea\right)
\eeq
and the mass-squared matrices $M_{H1}^2$ of
$H^c_1$, $\bar{H}^c_1$ and $M_{H2}^2$ of $H^c_2$,
$\bar{H}^c_2$
are
\beq
M_{H1(H2)}^{2}=\left(\ba{cc}
\gamma^2|S|^2  & \mp\gamma\ka M^2  \\
\mp\gamma\ka M^2 & \gamma^2|S|^2 \ea\right).
\eeq
The matrices $M_{\phi1(\phi2)}^2$ are always
positive definite, while the matrices
$M_{H1(H2)}^2$ acquire one negative eigenvalue
for
\beq
|S|<S_c\equiv\sqrt{\frac{\ka}{\gamma}}\;M.
\eeq
Thus, the trivial flat direction is now stable
for $|S|>S_c$ and unstable for $|S|<S_c$.
Yet, one can easily see that, for $\gamma\to 0$,
$S_c\to\infty$ and we are led to the previous
($\gamma=0$) case where the entire trivial flat
direction was a path of saddle points. So, one
can imagine that, for small enough values of the
parameter $\gamma$, the trivial flat direction,
after its destabilization at the critical point,
forks into four valleys of local or global minima
(for fixed $|S|$) of the potential in
Eq.~\eqref{eq:Fpotg}, which resemble the valleys
for new smooth hybrid inflation described above
in the $\gamma=0$ case.

\begin{figure}[tp]
\centering
\includegraphics[width=\linewidth]{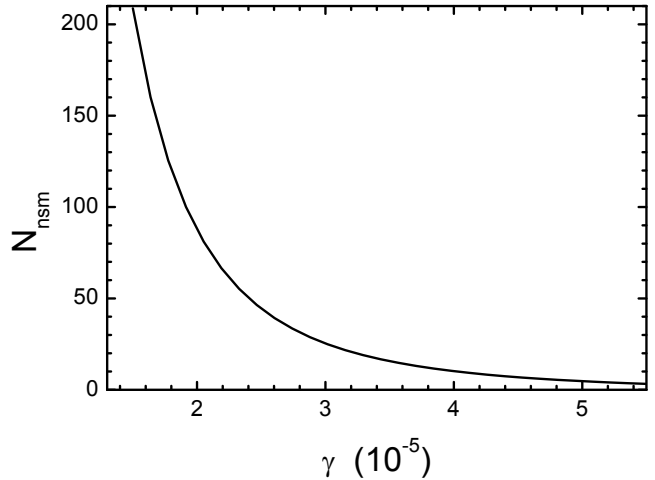}
\caption{Number of e-foldings $N_\text{nsm}$
along the new smooth inflationary path versus
$\gamma$ in global SUSY when the system slowly
rolls from $\sigma=0.95\,\sigma_c$ down to
$\sigma=\sigma_f$. The other parameters of the
model (except $\gamma$) take the values in
Eq.~\eqref{eq:parameters}.}
\label{fig:Nsmooth}
\end{figure}

\par
Actually, the valleys for a small non-zero
$\gamma$ are expected to differ from the ones for
$\gamma=0$ by corrections involving the small
parameter $\gamma$. The terms in the potential
of Eq.~\eqref{eq:Fpotg} which depend on $\gamma$
and the phases $\eps$, $\bar{\epsilon}$, and
$\theta$ are
\bea
\delta V&=&-2\gamma|H^c|^2\Big(\kappa M^2
\cos\theta-2\lambda|S||\pb|\cos\bar{\epsilon}
\nonumber\\
& &-\kappa|\phi|^2\cos(2\eps+\theta)\Big).
\label{eq:dV}
\eea
Estimating this expression on the valleys for
$\gamma=0$ by using the leading term in the
expansion of $|\phi|$ and $|\pb|$ in
Eq.~\eqref{eq:FieldExpansions}, we find that, for
$v_g/|S|<1$,
\bea
\delta V&\approx& -2\kappa\gamma M^2|H^c|^2
\Bigg(\cos\theta-\frac{2}{3}\cos\bar{\epsilon}
\nonumber\\
& &-\frac{1}{36}\left(\frac{v_g}{|S|}\right)^4
\cos(2\eps+\theta)\Bigg).
\label{eq:dVvalley}
\eea
From this, we see that the $\gamma$ dependent
corrections enhance the potential in the valleys
with $\eps=\bar{\epsilon}=\theta=\pi$ and reduce
it in the valleys with $\eps=\bar{\epsilon}=
\theta=0$. This fact can also be confirmed
numerically. So, as it turns out, the trivial
flat direction bifurcates at $|S|=S_c$ into two
valleys of {\em absolute} minima for fixed $|S|$
which correspond to $\theta\simeq 0$ and lead to
the two SUSY vacua in Eq.~\eqref{eq:vacuum1}.
They are the valleys for new smooth hybrid
inflation in the case with $\gamma\neq 0$, but
small. We should recall, however, that these two
valleys are not discrete, but belong to a
continuum of valleys.

\par
Unfortunately, it is quite difficult to find a
reliable expansion for the fields on these
valleys, mainly because of the obstacle at
$|S|=S_c$, which prevents us from taking the
limit $v_g/|S|\to 0$. So, numerical computation
is our last resort. We have found numerically
that, when the system crosses the critical point
at $\sigma=\sigma_c$ ($\sigma_c\equiv
\sqrt{2}S_c$) after it has rolled down the
trivial flat direction, does not immediately
settle down on the new smooth path. This takes
place after a while and at a value of $\sigma$
which is well above $0.95\,\sigma_c$.
Furthermore, quantum fluctuations which could
kick the system out of the new smooth path are
utterly suppressed well before the system reaches
this value of $\sigma$. However, to be on the
safe side, we will consider here the slow
rolling of the system along the new smooth path
starting from $\sigma=0.95\,\sigma_c$. In
Fig.~\ref{fig:Nsmooth}, we plot the number of
e-foldings $N_\text{nsm}$ along the new smooth
path as a function of the parameter $\gamma$ in
global SUSY and with the parameter values in
Eq.~\eqref{eq:parameters} when the system slowly
rolls from $\sigma=0.95\,\sigma_c$ down to
$\sigma=\sigma_f$, where $\eta=-1$ and the slow
rolling ends. We see that, for small enough
$\gamma$, we can have an adequate number of
e-foldings for solving the horizon and flatness
problems of standard hot big bang cosmology.

\begin{figure}[tp]
\centering
\includegraphics[width=\linewidth]{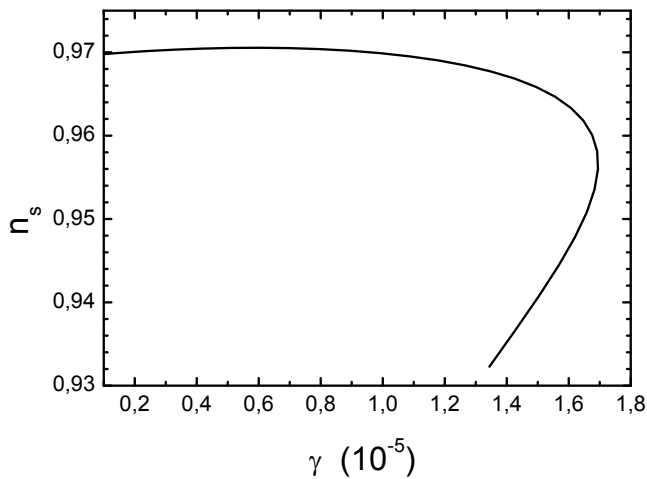}
\caption{Spectral index in new smooth hybrid
inflation versus $\gamma$ in global SUSY for
$p=\sqrt{2}\ka M/m=1/\sqrt{2}$ and $\ka=0.1$.
The endpoint of the curve at $n_s\simeq 0.932$
corresponds to the case where our present
horizon scale crosses outside the inflationary
horizon when $\sigma=0.95\,\sigma_c$.}
\label{fig:nsNewSmooth}
\end{figure}

\par
To pursue the investigation of the model further,
we set $p=\sqrt{2}\ka M/m=1/\sqrt{2}$, $\ka=0.1$
and fix the value of the power spectrum
$P_{\mathcal R}$ of the primordial curvature
perturbation to the three-year WMAP \cite{wmap3}
result $P_{\mathcal R}^{1/2}\simeq 4.85\ten{-5}$.
As already mentioned, on the new smooth path,
the fields
$H^c$ and  $\Hb^{c*}$ have practically the same
phase $(\theta\simeq 0)$. So, one of the vacua in
Eq.~\eqref{eq:vacuum1} is already selected during
inflation (i.e. the common phase of $H^c$ and
$\Hb^{c*}$ is fixed during inflation). We
set the VEV $|\vev{H^c}|=\sqrt{m\phi_+/\lambda}$
equal to the SUSY GUT scale (in practice, we just
put $v_g=\sqrt{mM/\lambda}\simeq 2.86\ten{16}
\units{GeV}$, since the resulting error is very
small). After these
choices, the only freedom left is the value of
$\gamma$. In Fig.~\ref{fig:nsNewSmooth}, we plot
the predicted spectral index of the model as a
function of $\gamma$. We terminate the curve when
the value of $\sigma$ at which our present
horizon crosses outside the inflationary horizon
becomes as large as $0.95\,\sigma_c$. We
observe that there exists a range of values for
$\gamma$ within which the system admits two
separate solutions, each corresponding to a
different value of $\lambda$. This new feature of
the model, which is not shared by conventional
smooth hybrid inflation, originates from the
presence of the critical point at
$\sigma=\sigma_c$ blocking the extension of the
new smooth path to larger values of $\sigma$. The
part of the curve with $n_s<0.96$ corresponds to
values of $\sigma_Q$ in the range
$0.85<\sigma_Q/\sigma_c<0.946$, while its branch
with $n_s>0.96$ corresponds to
$\sigma_Q/\sigma_c<0.85$. We see that spectral
indices compatible with Eq.~\eqref{eq:nswmap} can
easily be obtained for $\gamma$'s which are small
enough so that the number of e-foldings generated
is adequately large for solving the horizon and
flatness problems. It is important to point out
that, in global SUSY, the new smooth hybrid
inflation model is far superior to conventional
smooth hybrid inflation, which predicts
\cite{smooth} $n_s\simeq 0.969$, in that it can
easily accommodate much smaller values of $n_s$
and, thus, be more comfortably compatible with
data. However, we should note that obtaining
values of $n_s$ which are very close to its lower
bound in Eq.~(\ref{eq:nswmap}) would require
getting slightly above $\sigma=0.95\,\sigma_c$,
which is not impossible at all as we already
explained.

\par
For the values of $\gamma$ which correspond to
the curve depicted in Fig.~\ref{fig:nsNewSmooth},
i.e. $\gamma\simeq (0.3-1.7)\ten{-5}$, we find
that $\lambda\simeq (1.4-3.1)\ten{-3}$,
$M\simeq (2.4-3.6)\ten{16}\units{GeV}$, $m\simeq
(4.8-9.2)\ten{15}\units{GeV}$, and $\si_c\simeq
(3-9) \ten{17}\units{GeV}$. The number of
e-foldings from the time when the pivot scale
$k_0$ crosses outside the inflationary horizon
until the end of inflation is $N_Q\simeq
53.6-53.85$. The value $\sigma_f$ of $\sigma$
when inflation ends is about
$1.4\ten{17}\units{GeV}$ and $\si_Q$ lies in the
range $(2.85-3.025)\ten{17}\units{GeV}$. Finally,
$dn_s/d\ln k\simeq -(4.1-5.5)\ten{-4}$ and
$r\simeq (3-13)\ten{-7}$. Variations in the
values of $p$ and $\ka$ (which are the only
arbitrarily chosen parameters) have shown not to
have any significant effect on the results.
Contrary to the $\gamma=0$ case, the numerical
results for $\gamma\neq 0$ certainly depend on
the choice of the phases of the parameters in the
superpotential of Eq.~\eqref{eq:superpotential}.
As already explained, only one of the
dimensionless parameters of this superpotential,
say the parameter $\gamma$, is genuinely complex.
Its phase affects the position of the SUSY vacua
in Eq.~\eqref{eq:vacuum1} and presumably the
position of the new smooth paths which lead to
these vacua. However, the general qualitative
structure of the theory is not expected to
be affected.

\section{Supergravity corrections}
\label{sec:sugra}

\par
Following
Refs.~\cite{Bastero-Gil:2006cm,urRehman:2006hu},
we will show that when global SUSY is
promoted to local, some features of the model
are sensitive to non-minimal terms in the
K\"{a}hler potential. In particular, although
SUGRA corrections with a minimal K\"{a}hler
potential raise the spectral index above the
allowed range, non-minimal terms can help us to
reduce the spectral index so as to
become comfortably compatible with the data.
Once again, we will first concentrate on the case
$\gamma=0$ and then extend the model to include a
small, but non-zero $\gamma$.

\par
The F-term scalar potential in SUGRA is given by
\beq\label{eq:VSUGRA}
V=e^{K/m_{\rm P}^2}
\left[(F_i)^* K^{i^*j}F_j-3\,\frac{|W|^2}
{m_{\rm P}^2}\right],
\eeq
where $K$ is the K\"{a}hler potential,
$F_i=W_i+K_iW/m_{\rm P}^2$, a subscript $i$
($i^*$) denotes derivation with respect to the
complex scalar field $s^i$ ($s^{i{\,*}}$) and
$K^{i^*j}$ is the inverse of the K\"{a}hler
metric $K_{j\,i^*}$.

\par
We will consider, at first, a minimal K\"{a}hler
potential and leave the inclusion of non-minimal
terms for later. The minimal K\"{a}hler
potential, in our case, has the form
\beq\label{eq:MinKahler}
K_0=|S|^2+|\phi|^2+|\pb|^2+|H^c|^2+|\Hb^c|^2
\eeq
and the scalar potential is given by
\beq\label{eq:SUGRAmin}
\tilde{V}_0\equiv\frac{V_0}{\ka^2M^4}=
e^{K_0/m_{\rm P}^2}\;\left[\sum_{s}
\left|\tilde{W}_s+\frac{\tilde{W}s^*}
{m_{\rm P}^2}\right|^2-3\,\frac{|\tilde{W}|^2}
{m_{\rm P}^2}\right],
\eeq
where $\tilde{W}=W/\ka M^2$ and $s$ stands for
any of the five complex scalar fields appearing
in Eq.~\eqref{eq:MinKahler}. We have verified
numerically that, for the parameters in
Eq.~\eqref{eq:parameters} and $\gamma=0$, the
potential is again minimized for fixed $|S|$ on
the new smooth path with $\phi=\pm|\phi|$,
$\pb=\pm|\pb|$ and $H^c\Hb^c=\pm|H^c|^2$, where
the signs are correlated (recall that $S$ is
chosen real and positive). So, we will restrict
our attention again to these directions.
Furthermore, we have found that the relative
error in approximating the new smooth path by
Eq.~\eqref{eq:VarExpansions} or
\eqref{eq:FieldExpansions} is of the same order
of magnitude as that in the global SUSY limit
(see Fig.~\ref{fig:RelativeError}), namely $\sim
1\text{\textperthousand}$. So, we will use again
these expansions for the new smooth path in
SUGRA.

\par
Below, we give the expansions of the various
quantities entering the potential of
Eq.~\eqref{eq:SUGRAmin} calculated on the new
smooth path for $\gamma=0$. Note that, besides
$w$, we now have another small variable, namely
$|S|/m_{\rm P}$, which is expected to be at least
one order of magnitude below unity during
inflation (e.g. $S_Q/m_{\rm P}\sim 0.08$ for the
relevant value of $v_g$). In addition, the
constants
$v_g/m_{\rm P}$ and $M/m_{\rm P}$ are also well
below unity. We will treat only $v_g/m_{\rm P}$
as an independent small constant since
$M/m_{\rm P}=M/v_g\cdot v_g/m_{\rm P}$ with
$M/v_g \sim 1$. Using
Eqs.~\eqref{eq:VarExpansions} and
\eqref{eq:xyz1}-\eqref{eq:xyz}, we can expand
the superpotential and its derivatives on the new
smooth path as follows:
\bea
\frac{\tilde{W}}{m_{\rm P}} &\simeq& \frac{|S|}
{m_{\rm P}}\Big[1-\frac{1}{2}\,x_2(1-4x_2)\,w^4+
\dots\Big],\label{eq:WExpansion}\\
\tilde{W}_S &\simeq& \Big[1-x_2^2\,w^4+\dots
\Big],\label{eq:WsExpansion}\\
\tilde{W}_{\phi} &\simeq& \pm\Big[-4x_2z_2
\frac{M}{v_g}\,w^3+\dots\Big],
\label{eq:WphiExpansion}\\
\tilde{W}_{\pb} &\simeq& \pm\Big[2
\sqrt{2}p\,x_2^2\,w^2+\dots\Big],
\label{eq:WpbExpansion}\\
\tilde{W}_{H^c} &=& \pm\tilde{W}_{\Hb^c} \simeq
\Big[-2x_2\sqrt{z_2}\,w^2+\dots\Big],
\label{eq:WhExpansion}
\eea
where the $\pm$ signs are again correlated, the
ellipses represent terms of higher order in $w$, and
Eq.~\eqref{eq:WhExpansion} has been written in
the case where $H^c>0$ (for $H^c<0$, we should
put an overall minus sign in front of the
bracket). Using Eq.~\eqref{eq:VarExpansions}, we
can write the expansions of the fields on the new
smooth path as
\bea
\frac{\phi}{m_{\rm P}}&\simeq&\pm\frac{M}
{m_{\rm P}}\Big[x_2w^2+x_4w^4+\dots\Big],
\label{eq:phexp}\\
\frac{\pb}{m_{\rm P}}&\simeq&\pm\sqrt{2}p
\frac{v_g}{m_{\rm P}}\Big[y_1w+y_3w^3+
\dots\Big],\label{eq:pbexp}\\
\frac{H^c}{m_{\rm P}}&=&\pm\frac{\bar{H}^c}
{m_{\rm P}}\simeq\frac{v_g}{m_{\rm P}}
\Big[\sqrt{z_2}w+\frac{z_4}{2\sqrt{z_2}}w^3
+\dots\Big],\label{eq:hexp}
\eea
where the $\pm$ signs are correlated with the
ones in
Eqs.~\eqref{eq:WphiExpansion}-\eqref{eq:WhExpansion}
and we again take the case $H^c>0$.

\par
We will seek an expansion of the dimensionless
potential $\tilde{V}_0$ on the new smooth path
(for $\gamma=0$) in powers of $|S|/m_{\rm P}$ and
$w$. One can easily show, using
Eqs.~\eqref{eq:SUGRAmin}-\eqref{eq:hexp},
that only even powers of $|S|/m_{\rm P}$ and $w$
enter this expansion. Thus, the dimensionless
potential expanded in these variables up to
fourth order takes the form
\beq\label{eq:VSUGRAguess}
\tilde{V}_0\simeq A_0+A_2\frac{|S|^2}
{m_{\rm P}^2}+A_4\frac{|S|^4}{m_{\rm P}^4}
+B_2w^2+B_4w^4.
\eeq
To construct the expansion of the dimensionless
potential on the new smooth inflationary path, we
first classify the various possible types of
dimensionless quantities entering the calculation
of $\tilde{V}_0$ on this path. The dimensionless
parameters $p$, $x_i$, $y_i$, $z_i$, $\la/\ka$,
and $M/v_g$ will be considered to be of order
unity and, as all the quantities of order unity,
will be called of type $\1$. Any quantity that is
proportional to some positive power of
$w=v_g/|S|$ with coefficient of order unity will
be called of type $\ti$. Note that all the terms
in the square brackets in
Eqs.~\eqref{eq:WExpansion}-\eqref{eq:hexp}
are either of type $\1$ or $\ti$. Furthermore,
any quantity that is proportional to some positive
power of $|S|/m_{\rm P}$ with coefficient of order
unity will be called of type $\tii$. Finally,
positive powers of the small constant
$v_g/m_{\rm P}$ with coefficients of order unity
will be called quantities of type $\m$. It is
easy to see, using
Eqs.~\eqref{eq:SUGRAmin}-\eqref{eq:hexp}, that
only even powers of $v_g/m_{\rm P}$ appear in the
expansion of $\tilde{V}_0$. Quantities of the
form $\ti\cdot\tii$ can only take one of the
forms $\m$, $\m\cdot\ti$, and $\m\cdot\tii$. So,
the final expansion of $\tilde{V}_0$ is expected
to contain only terms of the form $\1$, $\ti$,
$\tii$, $\m$, $\m\cdot\ti$, and $\m\cdot\tii$.

\par
Now, we can split the relevant range $v_g\lsim
|S|\lsim m_{\rm P}$ of $|S|$ into two intervals
according to which of the two fourth order
quantities $v_g^4/|S|^4$ and $|S|^4/m_{\rm P}^4$
dominates. The former dominates in the interval
$v_g\lsim |S|\lsim (v_g m_{\rm P})^{1/2}$, while
the latter in the interval $(v_g m_{\rm P})^{1/2}
\lsim |S|\lsim m_{\rm P}$. Comparing the quantity
$v_g^2/m_{\rm P}^2$ with the two aforementioned
fourth order quantities, we find that, in each
of the two intervals, it is smaller than the
dominant fourth order quantity in this interval.
So, all the terms of type $\m$ can be neglected in
the final expression of the potential in
Eq.~\eqref{eq:VSUGRAguess} provided that $A_4$
and $B_4$ contain terms of type $\1$, which turns
out to be the case (see below). The same is true
for the terms of order $v_g^2/m_{\rm P}^2
\cdot v_g^2/|S|^2$ and $v_g^2/m_{\rm P}^2\cdot
|S|^2/m_{\rm P}^2$ as well as all the higher
order terms of the form $\m\cdot\ti$ and $\m\cdot
\tii$. According to the above, the dimensionless
potential to fourth order in $|S|/m_{\rm P}$ and
$w$ should only contain terms of type $\1$,
$\ti$, and $\tii$, which is equivalent to saying
that the coefficients $A_i$ and $B_i$ in
Eq.~\eqref{eq:VSUGRAguess} should not contain
terms of type $\m$.

\par
Let us now find some rules which can help us
manipulate the expansion of $\tilde{V}_0$ on the
new smooth path. First of all, note that this
dimensionless potential consists of a sum of
products of $\tilde{W}/m_{\rm P}$, $\tilde{W}_s$,
and $|s|/m_{\rm P}$, as seen from
Eq.~\eqref{eq:SUGRAmin}. The quantities
$|s|/m_{\rm P}$ with $s\ne S$ in
Eqs.~\eqref{eq:phexp}-\eqref{eq:hexp} consist
of terms of the form $\m\cdot\ti$, while
$|S|/m_{\rm P}$ and the quantities in
Eqs.~\eqref{eq:WExpansion}-\eqref{eq:WhExpansion}
contain terms of the form $\1$, $\ti$, $\tii$,
and $\m\cdot\ti$. It is readily shown that
products of any
of these quantities can only contain terms of
type $\1$, $\ti$, $\tii$, $\m$, $\m\cdot\ti$, and
$\m\cdot\tii$. Moreover, one can easily see that,
if a term of type $\m$, $\m\cdot\ti$, or
$\m\cdot\tii$ is encountered at any intermediate
stage of the calculation, it is bound to yield
terms of type $\m$, $\m\cdot\ti$, or
$\m\cdot\tii$ in the final expansion of
$\tilde{V}_0$. However, we have already shown
that such terms should not be kept in the final
form of the potential since they give a
negligible contribution. Thus, we conclude that
we can drop terms of the form $\m$, $\m\cdot\ti$,
and $\m\cdot\tii$ whenever we come across them
and maintain only terms of the form $\1$, $\ti$,
and $\tii$ in the various stages of the
calculation. A corollary to this is that we
can take $K_0$ in the exponential of
Eq.~\eqref{eq:SUGRAmin} to be simply $|S|^2$ and
$\tilde{W}/m_{\rm P}$ in
Eq.~\eqref{eq:WExpansion} to be simply
$|S|/m_{\rm P}$.

\par
Taking all the above into account, we can now
quite easily find that the relevant terms in the
dimensionless potential of
Eq.~\eqref{eq:WExpansion} on the new smooth path
(for $\gamma=0$) will be all contained in
\bea
\tilde{V}_0&\simeq& e^{|S|^2/m_{\rm P}^2}\;\Bigg[
\tilde{V}_g+\frac{|\tilde{W}|^2|S|^2}
{m_{\rm P}^4}\nonumber\\
& &+\left(\frac{\tilde{W}_S^*\tilde{W}S^*}
{m_{\rm P}^2}+\text{c.c}\right)
-3\,\frac{|\tilde{W}|^2}{m_{\rm P}^2}\Bigg],
\label{eq:tV0}
\eea
where $\tilde{V}_g=\sum_s|\tilde{W}_s|^2$ is the
dimensionless scalar potential in the global SUSY
limit. Substituting
Eqs.~\eqref{eq:WExpansion}-\eqref{eq:WhExpansion}
into Eq.~\eqref{eq:tV0} and keeping only the
relevant terms, we obtain the potential
\beq\label{eq:Vminsugra}
V_0\simeq\ka^2M^4\left(
1+\frac{1}{2}\,\frac{|S|^4}{m_{\rm P}^4}
-\frac{v_g^4}{54|S|^4}\right).
\eeq
Note that, in our case, the leading SUGRA
correction to the inflationary potential for
minimal K\"{a}hler potential, which corresponds
to the second term in the parenthesis in the
right hand side of Eq.~(\ref{eq:Vminsugra}), is
the same as the one found in Ref.~\cite{cop} in
the case of standard hybrid inflation and in
Ref.~\cite{senoguz} in the case of shifted and
smooth hybrid inflation. Actually, the
inflationary potential for conventional smooth
hybrid inflation in Ref.~\cite{senoguz} coincides
with the potential in Eq.~(\ref{eq:Vminsugra}),
which applies to new smooth hybrid inflation for
$\gamma=0$.

\par
Let us now turn to the consideration of a more
general K\"{a}hler potential containing
non-minimal
terms. As we are interested in the region of
field space with $|s|\ll m_{\rm P}$, we can
expand the K\"{a}hler potential as a power series
in the fields. The same rules that we have
extracted above for manipulating the expansion of
the potential on the new smooth path in the case
of minimal K\"{a}hler potential hold for this case
as well. In particular, in expanding the
potential up to
fourth order in $|S|/m_{\rm P}$ and $w$, we can
drop terms of the form $\m$, $\m\cdot\ti$, and
$\m\cdot\tii$ whenever they appear at an
intermediate stage of the calculation. As a
consequence, we can take $K$ in the exponential
of Eq.~\eqref{eq:VSUGRA} to consist only of terms
containing solely powers of the field $S$ and not
the other fields (compare with the similar
argument above in the case of a minimal
K\"{a}hler potential). Since terms of the form
$|S|^n(S^m+S^{*m})$ with $n\geq 0$ and $m\geq 1$
are not allowed due to the R symmetry, the only
relevant non-minimal K\"{a}hler potential terms
are
\beq\label{eq:KahlerTerms1}
|S|^4/m_{\rm P}^2,\quad |S|^6/m_{\rm P}^4
\eeq
up to order six in $|S|/m_{\rm P}$. The same
terms are the only non-minimal K\"{a}hler
potential
terms (up to sixth order) which can give a
non-negligible contribution to $K_i/m_{\rm P}$.
This is due to the fact that, in $K$, we cannot
have terms with a single field $s\ne S$
multiplying powers of $S$ and $S^*$ since there
exist no other gauge singlet fields in the
theory. So, all terms in $K$ other than the ones
of the form in Eq.~\eqref{eq:KahlerTerms1}
contain at least two fields $s\neq S$ and, thus,
give negligible contributions to $K_i/m_{\rm P}$.
Finally, the inverse K\"{a}hler metric $K^{i^*j}$
can be expanded as a power series of the higher
order terms contained in the K\"{a}hler metric
$K_{j\,i^*}$. Besides the terms of the form in
Eq.~\eqref{eq:KahlerTerms1}, other K\"{a}hler
potential terms that can contribute to
$K_{j\,i^*}$ are certainly the ones of the form
\beq\label{eq:KahlerTerms2}
|S|^2|s|^2/m_{\rm P}^2
\eeq
with $s$ being any of the fields $\phi$, $\pb$,
$H^c$, and $\Hb^c$. In general, any terms
containing two of the four fields $\phi$, $\pb$,
$H^c$, and $\Hb^c$ multiplied by powers of $S$
and $S^*$ will contribute. The only possible
combinations of two fields $s\neq S$ other than
$|s|^2$ that respect gauge invariance are
$H^c\Hb^c$, $\phi^2$, $\phi\pb$, $\phi^*\pb$, and
$\pb^2$ along with their complex conjugates. The
first two can be multiplied by powers of $|S|^2$,
while the other three need some extra $S$ or
$S^*$ factors in order to become R symmetry
invariant. In summary, we can parameterize the
most general K\"{a}hler potential which is
relevant for our calculation here as follows:
\bea
\label{eq:GeneralKahler}
K&=&K_0+\frac{k_S}{4}\,\frac{|S|^4}{m_{\rm P}^2}+
\frac{k_{SS}}{6}\,\frac{|S|^6}{m_{\rm P}^4}
+\sum_{s\ne S}k_{Ss}\,\frac{|S|^2|s|^2}
{m_{\rm P}^2}\nonumber\\
& &+\Bigg(k_{\phi\pb S^*}\frac{\phi\pb S^*}
{m_{\rm P}}+ k_{\phi^*\pb S^*}\frac{\phi^*\pb S^*}
{m_{\rm P}}+k_{\pb\pb S^*S^*}\frac{\pb^2 S^{*2}}
{m_{\rm P}^2}\nonumber\\
& &+k_{\phi\phi SS^*}\frac{\phi^2 |S|^2}
{m_{\rm P}^2}+k_{H\Hb SS^*}\frac{H^c\Hb^c |S|^2}
{m_{\rm P}^2}+\text{c.c.}\Bigg),
\eea
where the various $k$ coefficients are considered
to be of order unity. From this, we get
\beq\label{eq:dKdS}
\frac{K_S}{m_{\rm P}}\simeq\frac{S^*}{m_{\rm P}}
\left(1+\frac{k_S}{2}\,\frac{|S|^2}{m_{\rm P}^2}
+\frac{k_{SS}}{2}\,\frac{|S|^4}{m_{\rm P}^4}
\right),
\eeq
while all the other first derivatives
$K_s/m_{\rm P}$ are of the form $\m$,
$\m\cdot\ti$, or $\m\cdot\tii$ and can be
neglected. The relevant contributions to the
K\"{a}hler metric and its inverse are
\beq
\big(K_{j\,i^*}\big)\simeq
\left(\ba{ccccc}
K_{11^*} & 0 & 0 & 0 & 0 \\
0 & K_{22^*} & K_{23^*} & 0 & 0 \\
0 & K_{32^*} & K_{33^*} & 0 & 0 \\
0 & 0 & 0 & K_{44^*} & 0 \\
0 & 0 & 0 & 0 & K_{55^*}
\ea\right),
\eeq
\beq\label{eq:InverseK}
\big(K^{i^*j}\,\big)\simeq
\left(\ba{ccccc}
\frac{1}{K_{11^*}} & 0 & 0 & 0 & 0 \\[3pt]
0 & \frac{K_{33^*}}{D} & -\frac{K_{23^*}}{D}
& 0 & 0 \\[3pt]
0 & -\frac{K_{32^*}}{D} & \frac{K_{22^*}}{D}
& 0 & 0 \\[3pt]
0 & 0 & 0 & \frac{1}{K_{44^*}} & 0 \\[3pt]
0 & 0 & 0 & 0 & \frac{1}{K_{55^*}}
\ea\right),
\eeq
where
\begin{gather}
K_{11^*}\simeq 1+k_S\frac{|S|^2}{m_{\rm P}^2}+
\frac{3}{2}\,k_{SS}\frac{|S|^4}{m_{\rm P}^4},
\label{eq:ddKdSdS}\\
K_{22^*} = 1+k_{S\phi}\frac{|S|^2}{m_{\rm P}^2},
\quad K_{33^*} = 1+k_{S\pb}\frac{|S|^2}
{m_{\rm P}^2},\\
K_{44^*} = 1+k_{SH}\frac{|S|^2}{m_{\rm P}^2},
\quad K_{55^*} = 1+k_{S\Hb}\frac{|S|^2}
{m_{\rm P}^2},\\
K_{23^*}=K_{32^*}^*=k^*_{\phi^*\pb S^*}\frac{S}
{m_{\rm P}},\\
D=K_{22^*}K_{33^*}-|K_{23^*}|^2,
\end{gather}
and $i=1,2,3,4,5$ correspond to the fields
$S$, $\phi$, $\pb$, $H^c$, $\bar{H}^c$
respectively.

\par
As can be seen from Eqs.~\eqref{eq:VSUGRA} and
\eqref{eq:InverseK}, the only contribution to the
scalar potential on the new smooth path
originating from non-diagonal elements of the
inverse K\"{a}hler metric comes from the term
$(F_2)^* K^{2^*3}F_3+\text{c.c}$, which, on the
new smooth path, can be approximated to leading
order by
\beq
\ka^2M^4\left(16\sqrt{2}p\,x_2^3z_2\,
\mbox{Re}\,k_{\phi^*\pb S^*}\right)
\frac{M}{m_{\rm P}}\;w^4.
\eeq
It is, thus, of the form $\m\cdot\ti$ and can be
dropped. From the diagonal entries in the inverse
K\"{a}hler metric, one finds that the relevant
contributions to the potential on the new smooth
path will come from
\bea
\label{eq:VDom}
V&\simeq& e^{K/m_{\rm P}^2}
\Bigg[\left|W_S+\frac{WK_S}
{m_{\rm P}^2}\right|^2K^{S^*S}
\nonumber\\
& &+\sum_{s\ne S}|W_s|^2-3\,\frac{|W|^2}
{m_{\rm P}^2}\Bigg].
\eea
Substituting
Eqs.~\eqref{eq:WExpansion}-\eqref{eq:WhExpansion},
\eqref{eq:dKdS}, \eqref{eq:InverseK}, and
\eqref{eq:ddKdSdS} into Eq.~\eqref{eq:VDom},
expanding in powers of $|S|/m_{\rm P}$, and
keeping only terms of type $\1$, $\ti$ and $\tii$,
we finally obtain, for the potential on the new
smooth path for $\gamma=0$ in SUGRA, the
approximation
\beq\label{eq:VSUGRAeff}
V\simeq v_0^4\left(
1-k_S\frac{|S|^2}{m_{\rm P}^2}+\frac{1}{2}\,
\gamma_S\frac{|S|^4}{m_{\rm P}^4}
-\frac{v_g^4}{54|S|^4}\right),
\eeq
where $v_0=\sqrt{k}M$ and $\gamma_S\equiv 1-
\frac{7}{2}\,k_S-3\,k_{SS}+2\,k_S^2$. We see
that, from
the variety of terms in the K\"{a}hler potential,
only those with coefficients $k_S$ and $k_{SS}$
contribute to the scalar potential on the new
smooth path expanded up to fourth order in
$|S|/m_{\rm P}$ and $v_g/|S|$. Note that
Eq.~(\ref{eq:VSUGRAeff}) coincides with the
corresponding result found in
Ref.~\cite{urRehman:2006hu} in the case of
conventional smooth hybrid inflation. Moreover,
the SUGRA correction to the inflationary
potential which corresponds to the second and
third terms in the parenthesis in the right hand
side of Eq.~(\ref{eq:VSUGRAeff}) coincides with
the SUGRA correction found in
Ref.~\cite{Bastero-Gil:2006cm} in the case of
standard hybrid inflation.

\par
All the above results hold as long as
Eq.~\eqref{eq:FieldExpansions} is a good
approximation to the new smooth path for
$\gamma=0$ in the case of a non-minimal K\"ahler
potential too. We have checked numerically that,
at least for values of the parameters close to
the ones in Eq.~\eqref{eq:parameters}, the
relative error in the fields on the new smooth
path remains smaller than $2\%$ for a general
K\"ahler potential (which can include more terms
besides the ones shown in
Eq.~\eqref{eq:GeneralKahler}) even when the
various $k$ coefficients are of order unity.

\par
As in Ref.~\cite{Bastero-Gil:2006cm}, the new
terms in the inflationary potential which
originate from the non-minimal terms in the
K\"{a}hler potential and are proportional to
$|S|^2$ and $|S|^4$ can give rise to a local
minimum at $|S|=|S|_{\min}$ and maximum at
$|S|=|S|_{\max}<|S|_{\min}$ of the potential on
the inflationary path. This means that, if the
system starts from a point
with $|S|>|S|_{\max}$, it can be trapped in the
local minimum of the potential. Nevertheless, as
in Ref.~\cite{urRehman:2006hu} where conventional
smooth hybrid inflation was considered, in the
case of new smooth hybrid inflation too, there
exists a range of values for $k_S$ where the
minimum-maximum of the inflationary potential
does not appear and the system can start its slow
rolling from any point on the inflationary path
without the danger of getting trapped.

\par
Let us find the condition for the inflationary
potential in Eq.~\eqref{eq:VSUGRAeff}, which
holds in the case $\gamma=0$, not to have the
``minimum-maximum'' problem. Using the
dimensionless real inflaton field $\hat{\si}
\equiv\si/m_{\rm P}$, this potential and
its derivative with respect to $\hat{\si}$ are
given by
\bea
\tilde{V} &\equiv& \frac{V}{v_0^4}\simeq
1-\frac{1}{2}\,k_S\,\hat{\si}^2+\frac{1}{8}\,
\gamma_S\,\hat{\si}^4
-\frac{2\hat{v}_g^4}{27\hat{\si}^4},\\
\frac{d\tilde{V}}{d\hat{\si}} &\equiv&\frac{1}
{v_0^4}\,\frac{dV}{d\hat{\si}}
\simeq -k_S\,\hat{\si}+\frac{1}{2}\,\gamma_S\,
\hat{\si}^3+\frac{8\hat{v}_g^4}{27\hat{\si}^5},
\eea
where $\hat{v}_g\equiv v_g/m_{\rm P}$ and
$\gamma_S$ is assumed positive. We can evade the
local maximum and minimum of the inflationary
potential if we require that
$d\tilde{V}/d\hat{\si}$ remains positive for any
$\hat{\si}>0$ so that this potential is a
monotonically increasing function of $\si$. This
gives the condition
\beq
f(\hat{\si})\equiv\hat{\si}^8-\frac{2k_S}
{\gamma_S}\,\hat{\si}^6
+\frac{16\hat{v}_g^4}{27\gamma_S}\gtrsim 0.
\eeq
For $k_S>0$, which is the interesting case as we
will soon see, the minimum of $f(\hat{\si})$ lies
at $\hat{\si}_1=\left(3k_S/2\gamma_S
\right)^{1/2}$, where $f(\hat{\si}_1)=-27k_S^4/16
\gamma_S^4+16\hat{v}_g^4/27\gamma_S$ and the
requirement that $f(\hat{\si}_1)\gtrsim 0$ yields
the restriction
\beq\label{eq:kmax}
k_S\lesssim k_S^{\max}\equiv\frac{4}{3\sqrt{3}}\;
\gamma_S^{3/4}\frac{v_g}{m_{\rm P}}.
\eeq
Note that, for $\gamma_S\sim 1$, this inequality
implies that $\hat{\si}_1<1$ and, thus, the
minimum of $f(\hat{\si})$ lies in the relevant
region where $\si<m_{\rm P}$.

\par
For $k_S\gtrsim k_S^{\max}$, on the other hand,
the inflationary potential has a local minimum
and maximum which approximately lie at
\beq
\si_{\min}\simeq m_{\rm P}\left(\frac{2k_S}
{\gamma_S}\right)^{1/2},\quad
\si_{\max}\simeq m_{\rm P}\left(\frac{8v_g^4}
{27k_Sm_{\rm P}^4}\right)^{1/6}.
\eeq
Even in this case, the system can always undergo
hybrid inflation with the required number of
e-foldings starting at a $\si<\si_{\max}$. This
is due to the vanishing of the derivative
$V^{(1)}$ at $\si=\si_{\max}$. However, the more
the e-foldings we want to obtain the closer we
must set the initial $\si$ to the maximum of the
potential, which leads to an initial condition
problem. Yet, as we will see, we can obtain a
spectral index as low as $0.95$ at $k_0=0.002
\units{Mpc}^{-1}$ in agreement with the WMAP
three-year value $0.958\pm0.016$ \cite{wmap3}
maintaining the constraint
$k_S\lesssim k_S^{\max}$.

\par
Using the inflationary potential in
Eq.~\eqref{eq:VSUGRAeff}, the spectral index of
density perturbations turns out to be
\beq
n_s\simeq 1+2\eta_Q\simeq
1-2k_S+3\,\gamma_S\,\frac{\si^2_Q}{m_{\rm P}^2}-
\frac{80v_g^4m_{\rm P}^2}{27\si^6_Q},
\eeq
where $\eta_Q$ is the value of $\eta$ when the
pivot scale $k_0=0.002\units{Mpc}^{-1}$ crosses
outside the inflationary horizon. We can see
that the $k_S$ term in the K\"{a}hler potential
contributes to the lowering of the spectral
index if $k_S$ is positive. So, a $k_S$ with this
choice of its sign can help us to make the
spectral index comfortably compatible with the
three-year WMAP measurements \cite{wmap3}.
However, since we cannot have any reliable and
convenient approximation for $\si_Q$, a numerical
investigation is required.

\begin{figure}[tp]
\centering
\includegraphics[width=\linewidth]{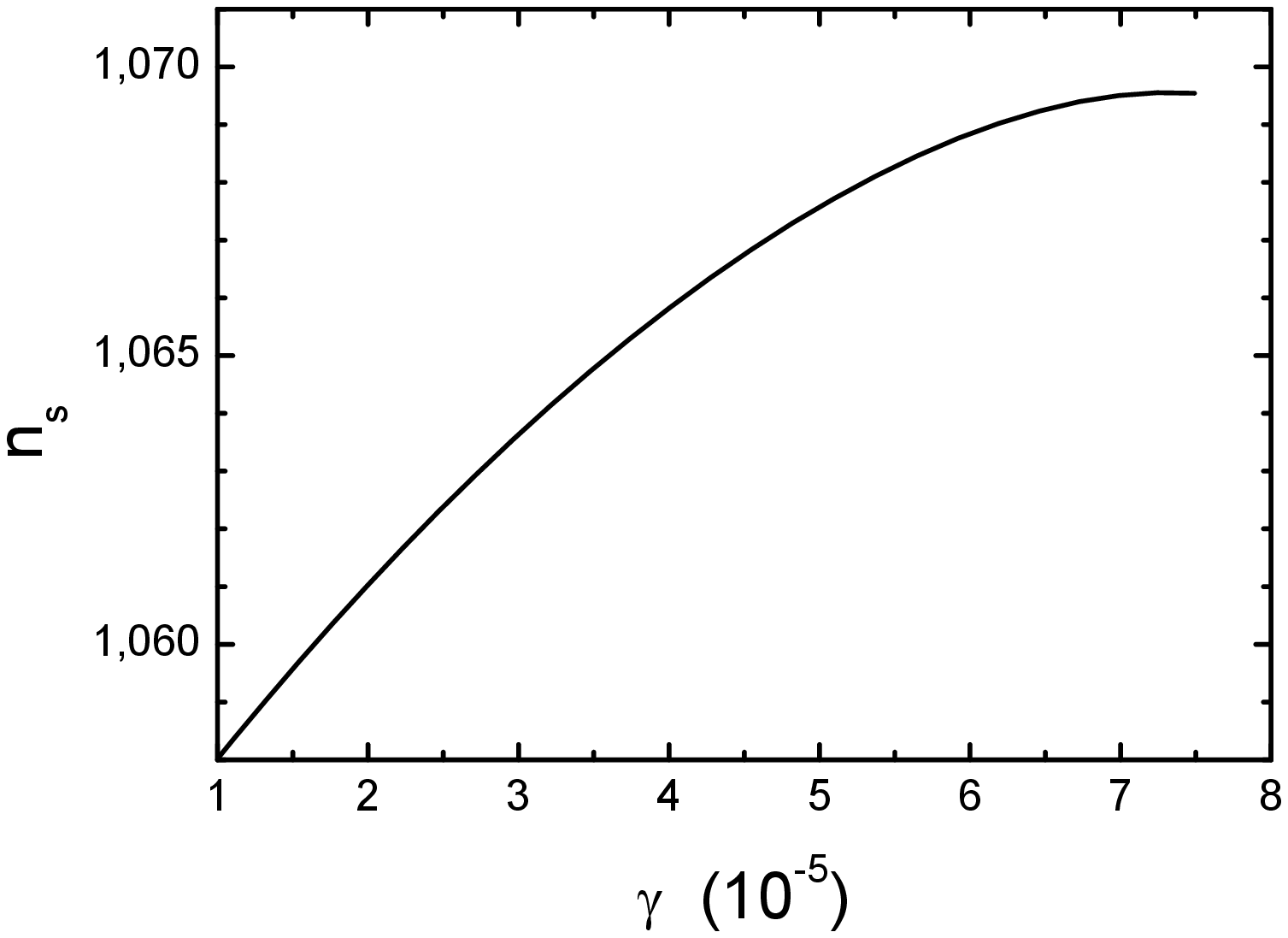}
\caption{Spectral index in new smooth hybrid
inflation versus $\gamma$ in minimal SUGRA for
$p=\sqrt{2}\ka M/m=1/\sqrt{2}$, $\ka=0.1$. The
endpoint of the curve at $\gamma\simeq 0.75
\ten{-6}$ ($n_s\simeq 1.0695$) corresponds to the
case where our present horizon scale crosses
outside the inflationary horizon when $\sigma=
0.95\,\sigma_c$.}
\label{fig:nsMinSUGRA}
\end{figure}

\par
Turning now to the case of small, but non-zero
$\gamma$, one can assert that again only the same
non-minimal terms of the K\"{a}hler potential
with
coefficients $k_S$ and $k_{SS}$ will enter the
expansion of the potential on the new smooth
path, although the global SUSY potential for new
smooth hybrid inflation is not, in this case,
given by Eq.~\eqref{eq:Vg0} but has to be
calculated numerically. So, due to
the small value of $\gamma$, we can assume that
the potential on the new smooth path in the case
of SUGRA with the non-minimal K\"{a}hler
potential of Eq.~(\ref{eq:GeneralKahler}) and
$\gamma\neq 0$ has the form
\beq
V\simeq v_0^4\left(\tilde{V}_\text{SUSY}
-\frac{1}{2}\,k_S\,\frac{\si^2}{m_{\rm P}^2}+
\frac{1}{8}\,\gamma_S\,\frac{\si^4}{m_{\rm P}^4}
\right),
\eeq
where $\tilde{V}_\text{SUSY}\equiv V_\text{SUSY}/
v_0^4$ with $V_\text{SUSY}$ being the
inflationary potential in the case of global SUSY
and $\gamma\neq0$. Note, also, that, in the
SUGRA and $\gamma\neq0$ case, the
critical value of $\sigma$, where the trivial flat
direction becomes unstable, will be slightly
different from the critical value of $\sigma$ in
the global SUSY case.

\par
As in the global SUSY case with $\gamma\neq 0$,
we take $p=\sqrt{2}\ka M/m=1/\sqrt{2}$,
$\ka=0.1$ and fix numerically the power spectrum
$P_{\mathcal R}$ of the primordial curvature
perturbation to the three-year WMAP normalization
\cite{wmap3} in SUGRA too. We
also set the VEV $|\vev{H^c}|$ equal to the SUSY
GUT scale, which, to a very good approximation,
means that we put $v_g=\sqrt{mM/\lambda}\simeq
2.86\ten{16}\units{GeV}$. The scalar spectral
index in SUGRA with a minimal K\"{a}hler
potential (i.e. $k_S=k_{SS}=0$) as a function of
the parameter $\gamma$
is shown in Fig.~\ref{fig:nsMinSUGRA}. We
terminate the curve when the value of $\sigma$
at which our present horizon scale crosses
outside the inflationary horizon reaches
$0.95\,\sigma_c$. We see that minimal SUGRA
elevates the scalar spectral index above the $95\%$
confidence level range obtained by fitting the
three-year WMAP data \cite{wmap3} by the standard
power-law $\Lambda$CDM cosmological model ($n_s$
tends to approximately $1.055$ as $\gamma\to0$).
This situation is readily rectified by the
inclusion of
non-minimal terms in the K\"{a}hler potential as
we will see below. For the range of values of
$\gamma$ shown in Fig.~\ref{fig:nsMinSUGRA} (i.e.
for $\gamma\sim(1-7.5)\ten{-5}$), the ranges of
the other parameters of the model are as follows:
$\lambda\simeq(1.33-1.68)\ten{-2}$,
$M\simeq(7.4-8.3)\ten{16}\units{GeV}$,
$m\simeq(1.48-1.66)\ten{16}\units{GeV}$,
$\sigma_c\simeq(4.2-9.8)\ten{17}\units{GeV}$,
$\sigma_Q\simeq(3.6-3.95)\ten{17}\units{GeV}$,
$\sigma_f\simeq(1.39-1.395)\ten{17}\units{GeV}$,
$N_Q\simeq54.3-54.4$,
$dn_s/d\ln k\simeq-(2.1-2.6)\ten{-3}$, and
$r\simeq(2.4-3.8)\ten{-5}$.

\begin{figure}[tp]
\centering
\includegraphics[width=\linewidth]{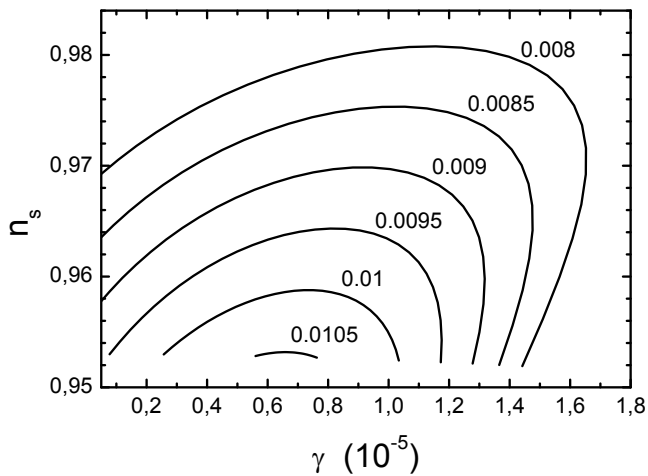}
\caption{Spectral index in new smooth hybrid
inflation in non-minimal SUGRA as a function of
$\gamma$ for $p=\sqrt{2}\ka M/m=1/\sqrt{2}$
and $\ka=0.1$. The values of $k_S$, which are
indicated on the curves, range from $0.008$ to
$0.0105$ and $k_{SS}=0$.}
\label{fig:SpectralIndex}
\end{figure}

\par
Next, we consider the case where non-minimal
terms are present in the K\"{a}hler potential. We
will let $k_S$ have a non-zero positive value,
but take $k_{SS}=0$ for simplicity. We calculate
numerically the spectral index and plot it in
Fig.~\ref{fig:SpectralIndex} as a function of
the parameter $\gamma$ for various values of
$k_{S}$. The limiting points on each curve
correspond to the situation where the potential
on the new smooth inflationary path starts
developing a local minimum and maximum. We
observe that, although all curves terminate on
the right, only curves that correspond to larger
values of $k_S$ (and smaller values of $n_s$)
have an endpoint on the small $\gamma$ side. It
is instructive to note that, for $\gamma=0$,
Eq.~\eqref{eq:kmax} gives $k_S^{\max}\simeq
0.0088$, which is in fairly good agreement with
Fig.~\ref{fig:SpectralIndex}. From this figure,
one can infer that the spectral index can be
readily set below unity in SUGRA with non-minimal
K\"{a}hler potential and that one can achieve a
value as low as $n_s\simeq 0.952$ without having
to put up with a local minimum and maximum of the
potential on the inflationary path. This minimal
value of $n_s$ corresponds to the endpoint of the
curve with $k_S=0.008$. The maximal allowed value
of $k_S$ is about 0.01054 corresponding to
$\gamma\simeq 0.66\ten{-5}$ and
$n_s\simeq 0.953$. Finally, for the range of
values of $\gamma$ and $k_S$ corresponding to the
curves in Fig.~\ref{fig:SpectralIndex}, the
ranges of variance of the other parameters of the
model are as follows:
$\lambda\simeq(1.5-2.6)\ten{-3}$,
$M\simeq(2.5-3.3)\ten{16}\units{GeV}$,
$m\simeq(0.5-0.66)\ten{16}\units{GeV}$,
$\sigma_c\simeq(0.45-1.7)\ten{18}\units{GeV}$,
$\sigma_Q\simeq(2.54-2.77)\ten{17}\units{GeV}$,
$\sigma_f\simeq(1.39-1.395)\ten{17}\units{GeV}$,
$N_Q\simeq53.6-53.8$,
$dn_s/d\ln k\simeq-(7.2-9.2)\ten{-4}$, and
$r\simeq(3-9.6)\ten{-7}$. Variations in the
values of $p$ and $\ka$ have shown not to have
any significant effect on the results. In
particular, the spectral index cannot become
smaller than about 0.95 by varying these
parameters provided that the appearance of a
local minimum and maximum of the inflationary
potential is avoided. Note, however, that smaller
values of $n_s$ can be readily achieved, but at
the cost of having the minimum-maximum problem.

\section{Conclusions}
\label{sec:conclusions}

\par
We considered the extension of the SUSY PS model
which has been introduced in Ref.~\cite{quasi} in
order to solve the $b$-quark mass problem in SUSY
GUT models with exact asymptotic Yukawa
unification, such as the simplest SUSY PS model,
and universal boundary conditions. This extended
model leads naturally to a (moderate) violation
of the asymptotic Yukawa unification so that, for
$\mu>0$, the predicted $b$-quark mass resides
within the experimentally allowed range.
Moreover, it is known that this model
automatically leads to the so-called new shifted
hybrid inflationary scenario, which is based only
on renormalizable superpotential terms and avoids
the cosmological disaster from a possible
overproduction of PS magnetic monopoles at the
end of inflation.

\par
Here, we have demonstrated that this PS model can
also lead to a new version of smooth hybrid
inflation, which, in contrast to the conventional
realization of smooth hybrid inflation, is based
only on renormalizable interactions. An important
prerequisite for this is that a particular
parameter of the superpotential is adequately
small. Then the scalar potential of the model
possesses, for a wide range of its other
parameters, valleys of minima with classical
inclination which can be used as inflationary
paths leading to a new realization of smooth
hybrid inflation. This scenario, in global SUSY,
is naturally consistent with the fitting of the
three-year WMAP data by the standard power-law
$\Lambda$CDM cosmological model. In particular,
the spectral index turns out to be adequately
small so that it is compatible with the data.
Moreover, as in the conventional realization of
smooth hybrid inflation, the PS gauge group is
already broken to the SM gauge group during new
smooth hybrid inflation and, thus, no
topological defects are formed at the end of
inflation. Therefore, the problem of possible
overproduction of PS magnetic monopoles is
solved.

\par
Embedding the model in SUGRA with a minimal
K\"{a}hler potential raises the scalar spectral
index to values which are too high to be
compatible with the recent data. However,
inclusion of a non-minimal term in the
K\"{a}hler potential with appropriately chosen
sign can help to reduce the spectral index so
that it resides comfortably within the allowed
range. The potential along the new smooth
inflationary path, however, can remain everywhere
a monotonically increasing function of the
inflaton field. So, unnatural restrictions on
the initial conditions for inflation due to the
appearance of a maximum and a minimum of the
potential on the new smooth inflationary path
when such a non-minimal K\"{a}hler potential is
used can be avoided.

\section*{ACKNOWLEDGEMENTS}

\par
This work was supported in part by the European
Commission under the Research and Training
Network contracts MRTN-CT-2004-503369 and
HPRN-CT-2006-035863. It was also supported in
part by the Greek Ministry of Education and
Religion and the EPEAK program Pythagoras.

\def\ijmp#1#2#3{{Int. Jour. Mod. Phys.}
{\bf #1},~#3~(#2)}
\def\plb#1#2#3{{Phys. Lett. B }{\bf #1},~#3~(#2)}
\def\zpc#1#2#3{{Z. Phys. C }{\bf #1},~#3~(#2)}
\def\prl#1#2#3{{Phys. Rev. Lett.}
{\bf #1},~#3~(#2)}
\def\rmp#1#2#3{{Rev. Mod. Phys.}
{\bf #1},~#3~(#2)}
\def\prep#1#2#3{{Phys. Rep. }{\bf #1},~#3~(#2)}
\def\prd#1#2#3{{Phys. Rev. D }{\bf #1},~#3~(#2)}
\def\npb#1#2#3{{Nucl. Phys. }{\bf B#1},~#3~(#2)}
\def\npps#1#2#3{{Nucl. Phys. B (Proc. Sup.)}
{\bf #1},~#3~(#2)}
\def\mpl#1#2#3{{Mod. Phys. Lett.}
{\bf #1},~#3~(#2)}
\def\arnps#1#2#3{{Annu. Rev. Nucl. Part. Sci.}
{\bf #1},~#3~(#2)}
\def\sjnp#1#2#3{{Sov. J. Nucl. Phys.}
{\bf #1},~#3~(#2)}
\def\jetp#1#2#3{{JETP Lett. }{\bf #1},~#3~(#2)}
\def\app#1#2#3{{Acta Phys. Polon.}
{\bf #1},~#3~(#2)}
\def\rnc#1#2#3{{Riv. Nuovo Cim.}
{\bf #1},~#3~(#2)}
\def\ap#1#2#3{{Ann. Phys. }{\bf #1},~#3~(#2)}
\def\ptp#1#2#3{{Prog. Theor. Phys.}
{\bf #1},~#3~(#2)}
\def\apjl#1#2#3{{Astrophys. J. Lett.}
{\bf #1},~#3~(#2)}
\def\n#1#2#3{{Nature }{\bf #1},~#3~(#2)}
\def\apj#1#2#3{{Astrophys. J.}
{\bf #1},~#3~(#2)}
\def\anj#1#2#3{{Astron. J. }{\bf #1},~#3~(#2)}
\def\apjs#1#2#3{{Astrophys. J. Suppl.}
{\bf #1},~#3~(#2)}
\def\mnras#1#2#3{{MNRAS }{\bf #1},~#3~(#2)}
\def\grg#1#2#3{{Gen. Rel. Grav.}
{\bf #1},~#3~(#2)}
\def\s#1#2#3{{Science }{\bf #1},~#3~(#2)}
\def\baas#1#2#3{{Bull. Am. Astron. Soc.}
{\bf #1},~#3~(#2)}
\def\ibid#1#2#3{{\it ibid. }{\bf #1},~#3~(#2)}
\def\cpc#1#2#3{{Comput. Phys. Commun.}
{\bf #1},~#3~(#2)}
\def\astp#1#2#3{{Astropart. Phys.}
{\bf #1},~#3~(#2)}
\def\epjc#1#2#3{{Eur. Phys. J. C}
{\bf #1},~#3~(#2)}
\def\nima#1#2#3{{Nucl. Instrum. Meth. A}
{\bf #1},~#3~(#2)}
\def\jhep#1#2#3{{J. High Energy Phys.}
{\bf #1},~#3~(#2)}
\def\lnp#1#2#3{{Lect. Notes Phys.}
{\bf #1},~#3~(#2)}
\def\appb#1#2#3{{Acta Phys. Polon. B}
{\bf #1},~#3~(#2)}
\def\njp#1#2#3{{New J. Phys.}
{\bf #1},~#3~(#2)}
\def\pl#1#2#3{{Phys. Lett. }{\bf #1B},~#3~(#2)}
\def\jcap#1#2#3{{J. Cosmol. Astropart. Phys.}
{\bf #1},~#3~(#2)}
\def\mpla#1#2#3{{Mod. Phys. Lett. A}
{\bf #1},~#3~(#2)}
\def\jpcs#1#2#3{{J. Phys. Conf. Ser.}
{\bf #1},~#3~(#2)}
\def\apjss#1#2#3{{Astrophys. J. Suppl. Ser.}
{\bf #1},~#3~(#2)}

\end{document}